\documentclass[prb,twocolumn]{revtex4}

\usepackage{amsmath}
\usepackage{graphicx}

\newcommand{\expt}[1]{\left< #1\right>}
\newcommand{\Eq}[1]{Eq.~(\ref{#1})}
\newcommand{\Eqs}[1]{Eqs.~(\ref{#1})}
\newcommand{\Fig}[1]{Fig.~\ref{#1}}
\newcommand{\Figs}[2]{Figs.~\ref{#1} and~\ref{#2}}
\newcommand{\Figure}[1]{Figure~\ref{#1}}
\newcommand{\Irms}{I_\mathrm{rms}}
\newcommand{\Jpara}{J_\parallel}
\newcommand{\Jperp}{J_\perp}

\newcommand{\teq}{\tau_\mathrm{eq}}
\newcommand{\tmax}{\tau_\mathrm{max}}
\newcommand{\taver}{\tau_\mathrm{aver}}
\newcommand{\Tabsim}{Tab.~\ref{tab:sets}}
\newcommand{\Tmin}{T_\mathrm{min}}
\newcommand{\Tmax}{T_\mathrm{max}}

\newcommand{\onlineRef}[1]{Ref.~\onlinecite{#1}}
\newcommand{\VG}{Ref.~\onlinecite{Olsson:vg-lett}}
\newcommand{\Uperp}{\Upsilon_\perp}
\newcommand{\Upara}{\Upsilon_\parallel}
\newcommand{\av}[1]{\left[#1\right]_\mathrm{av}}

\begin{document}
\title{Vortex glass transitions in disordered three-dimensional XY models:
  Simulations for several different sets of parameters}

\author{Peter Olsson}

\affiliation{Department of Physics, Ume\aa\ University, 
  901 87 Ume{\aa}, Sweden}

\date{\today}   

\begin{abstract}
  The anisotropic frustrated 3D XY model with strong disorder in the coupling
  constants is studied as a model of a disordered superconductor in an applied
  magnetic field.  Simulations with the exchange Monte Carlo method are
  performed for frustrations $f=1/5$ and $f=1/4$, corresponding to two different
  values of the magnetic field along the $z$ direction. The anisotropy is also
  varied. The determination of the helicity modulus from twist histograms is
  discussed in some detail and the helicity modulus is used in finite size
  scaling analyses of the vortex glass transition. The general picture is that
  the behavior in [Phys.\ Rev.\ Lett.\ \textbf{91}, 077002 (2003)] is confirmed.
  For strong (e.g.\ isotropic) coupling in the $z$ direction the helicity
  modulus fails to scale and it is argued that this is due to a too small
  effective randomness of such systems for the accessible system sizes.
\end{abstract}


\maketitle

\section{Introduction}

An applied magnetic field in a type-II superconductor will give rise to vortex
lines that penetrate the sample. A current applied perpendicular to these vortex
lines will give rise to a force perpendicular to both the current and the
magnetic field. In a pure system there is nothing that hinders the motion of the
vortex lines and their motion leads to flux-flow resistivity and therefore a
loss of superconductivity. The presence of point disorder could mean a
substantial reduction of the mobility of the vortex lines, but the resistivity
would, in the conventional picture, nevertheless always be non-zero.

A vortex glass phase is an alternative possibility that was suggested to restore
the true superconducting state.\cite{Fisher:89,Fisher_Fisher_Huse} The idea is
that the finite disorder strength together with the vortex line interaction
leads to diverging energy barriers against the vortex motion, and thereby a
vanishing resistivity.  This was suggested to take place through a continuous
transition with universal exponents and certain scaling properties.
Experimental results in support of this picture have been
reported,\cite{Gammel_SB,Klein_CMESSJ,Petrean_PKFC} but the conclusion of a
vortex glass phase has also often been
questioned.\cite{Reichhardt_Otterlo_Zimanyi}

There has also been much work on simulations of vortex glass models.  The
simplest three-dimensional (3D) vortex glass model, that was also the first to
be studied, is the 3D gauge glass model that includes the disorder through a
random vector potential added to the phase difference of the superconducting
order parameter. Already the early simulations\cite{Huse_Seung,Reger_TYF} found
strong evidence for a transition, and with the exchange Monte Carlo (MC)
technique\cite{Hukushima_Nemoto} it has been possible both to give more
convincing evidence for a transition and to determine the value of the
correlation length exponent to $\nu =1.39\pm0.04$.\cite{Katzgraber_Campbell}

A problem with using the 3D gauge glass as a model of a disordered
superconductor in an applied magnetic field, is the generally recognized fact
that the model lacks some of the properties and symmetries of the physical
system. The applied field both breaks the spatial symmetry of the system and
introduces an additional length scale. In a model that properly includes these
features one would e.g.\ have the possibility of anisotropic scaling, i.e.\ 
different divergences of the correlations parallel and perpendicular to the
applied field.

Several attempts have recently been done to simulate systems with the correct
symmetry. The first published results are from simulations of a frustrated 3D XY
model with filling $f=1/4$ and disorder in the coupling
constants.\cite{Kawamura:00} The correlation length exponent was there
determined to be $\nu=2.2$ even though the quality of the data did not allow for
any firm conclusions. In a second paper by the same author the open boundary
conditions employed in the first study were changed to standard periodic
boundary conditions.\cite{Kawamura:03} The data now rather suggested
$\nu\approx1.1$, but some quantities still failed to provide good scaling. Some
aspects of these simulations are discussed in Sec.~\ref{sec:discuss}.

Simulations have also been performed with vortex lines instead of the phase
variables of the XY model.\cite{Vestergren_Lidmar_Wallin} A simulation study of
such a vortex line model with strong point disorder gives the value $\nu=0.7$,
indistinguishable from the 3D XY exponent. In that study the pinning energy was
quite strong which presumably means that most plaquettes are either always
occupied or always empty. One possible reason for the 3D XY-like exponents could
then be that the model supports vortex loop excitations (as in the 3D XY model)
against a background of frozen-in field lines from the applied
field.\cite{Minnhagen}

The present paper is a sequel of \VG\ which gave the first numerical support for
3D gauge glass exponents in vortex glass simulations with the correct symmetry.
The approach was there to study an anisotropic model with much weaker couplings
in the field direction than in the directions perpendicular to the field. This
was a natural choice due to experiences from the first order transition between
the Abrikosov lattice and the vortex line liquid. In these
simulations\cite{Hu_M_Tachiki,Olsson_Teitel:xy3f} it has been found that the
correct behavior required a great flexibility of the field induced vortex lines,
which could be obtained either with a very large size of the system along the
direction of the applied field or with weaker couplings between the phase angles
in the same direction. As we will see below the choice of an anisotropic model
turns out to be crucial for obtaining convincing scaling collapses.

Another recent study of the vortex glass transition has been done on a model
that extends the elastic description of a vortex lattice to include
dislocations.\cite{Lidmar:vg} The correlation length exponent of the transition
was found to be $\nu \approx 1.3$, which within reasonable error bars also is
consistent with 3D gauge glass universality.

In the present paper we present detailed analyses of the frustrated 3D XY model
with strong disorder in the coupling constants. The paper is an extension of
\VG\ in two respects: (i) The determination of the helicity modulus from
simulations with twist fluctuations as well as analyses of the thermalization
and the exchange steps in the Monte Carlo simulations are described in
considerably more detail. (ii) Simulations and analyses have been done for
several different sets of parameters.

The organization of the paper is as follows: In Sec.\ II we discuss the
determination of the helicity modulus from twist histograms. Section III deals
with the vortex glass model and the different sets of parameters used in the
simulations. In Sec.\ IV the simulation methods are discussed with emphasis on
some aspects of the exchange Monte Carlo technique, and Sec.\ V gives the
simulation results. Section VI, finally, contains a discussion together with a
short summary.

\section{Determination of the helicity modulus}

The XY model is defined by the Hamiltonian
\begin{displaymath}
  H = \sum_{\langle ij\rangle} U(\theta_i - \theta_j),
\end{displaymath}
where a common choice for the spin interaction is $U(\phi)=-J\cos(\phi)$.
The helicity modulus, which is the standard probe of phase coherence in XY
models, is defined through the response to an applied twist. One way to define
the twist is to generalize the standard periodic boundary conditions
$\theta_{(L,0,0)} = \theta_{(0,0,0)}$ to
\begin{displaymath}
  \theta_{(L,0,0)} = \theta_{(0,0,0)} + \Delta_x,
\end{displaymath}
and similarly in the other directions. Here $\Delta_x$ is the phase mismatch or
the \emph{total twist} in the $x$ direction. One may alternatively think about
the twist as being spread out across the whole system and introduce the
\emph{twist per link}, $\delta_\mu = \Delta_\mu/L_\mu$. The Hamiltonian may then
be written
\begin{displaymath}
  H = \sum_i\sum_\mu U(\theta_{i+\mu} - \theta_i - \delta_\mu).
\end{displaymath}

The helicity modulus is defined through the change in the free energy
$F(\Delta_\mu)$, or the free energy per site $f=F/V$, as 
\begin{equation}
  \Upsilon_\mu = 
  \left.\frac{\partial^2 f}{\partial\delta_\mu^2}\right|_{\delta_\mu=0} = 
  \frac{L_\mu^2}{V}\left.\frac{\partial^2 F}{\partial \Delta_\mu^2}\right|_{\Delta_\mu=0},
  \label{d2FdDelta}
\end{equation}
which gives the correlation function\cite{Teitel_Jayaprakash:83a},
\begin{equation}
  \Upsilon_\mu = \frac{1}{V}\expt{\sum_i U''(\phi_{i\mu})} -
  \frac{1}{TV}\expt{\left(\sum_i U'(\phi_{i\mu})\right)^2}.
  \label{Ups-std}
\end{equation}
With this correlation function the determination of the helicity modulus is done
in simulations performed with zero twist.  Note that the derivative in
\Eq{Ups-std} is evaluated at the minimum of the free energy which typically is
$\Delta=0$. However, in some disordered models there is nothing that guarantees
that the minimum of the free energy is at zero twist. The approach taken here is
to study such systems with simulations that include the twist fluctuations as
additional dynamical variables.

\subsection{Twist fluctuations}

\begin{figure}[htbp]
  \includegraphics[width=8cm]{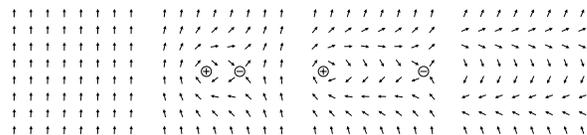}
  \caption{The separation of a vortex pair in a system with periodic
    boundary conditions gives a configuration with the phase rotating
    by $2\pi$.}
  \label{pair-pbc}
\end{figure}

\begin{figure}[htbp]
  \includegraphics[width=8cm]{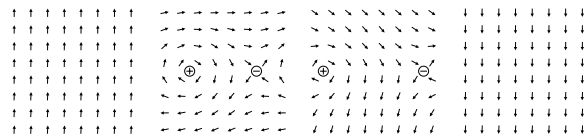}
  \caption{The separation of a vortex pair in a system with fluctuating twist
    boundary conditions. The twist variable is here applied between the top and
    the bottom rows of spins that are connected through the boundary conditions.
    The four panels are for $\Delta_y = 0$, $\pi$, $0.85\times 2\pi$, and
    $2\pi$, respectively.}
  \label{pair-ftbc}
\end{figure}

There is a well-known duality relation between an XY model in the Villain
representation and a gas of interacting charges. In two dimension this is a
Coulomb gas with logarithmic interactions and in three dimensions a gas of
interacting loops. As observed by several
authors\cite{Thijssen_Knops:88,Olsson:Vl,Vallat_Beck,Olsson:self-cons.long} the
XY model that is dual to a Coulomb gas with periodic boundary conditions also
includes twist fluctuations.  Physically, the twist fluctuations are necessary
for the process when a pair of vortices separate, cross the boundary and
recombine. In the absence of twist fluctuations such a process gives a
configuration where the phase rotates by $2\pi$ across the system in the
direction perpendicular to the vortex separation, as illustrated in
\Fig{pair-pbc}. The effect is that recombinations of vortices effectively is
prohibited.  \Figure{pair-ftbc} illustrates the vortex separation in the
presence of twist fluctuations in the $y$ direction.


\subsection{Basic relations}

An alternative means to obtain $\Upsilon_\mu$ is by first determining the free
energy, $F(\Delta_\mu)$. The simulations are then performed with fluctuating
twists in the $\mu$ direction and periodic boundary conditions in the other two
directions,
\begin{displaymath}
  H = \sum_{i,\lambda\neq\mu} U(\theta_{i+\lambda} - \theta_i)
  + \sum_{i} U(\theta_{i+\mu} - \theta_i + \Delta_\mu/L_\mu).
\end{displaymath}
The free energy is obtained from the histogram $P(\Delta_\mu)$ through
\begin{displaymath}
  F(\Delta_\mu) = -T\ln P(\Delta_\mu),
\end{displaymath}
and the helicity modulus may be determined from a fit of the free energy in a
narrow range $r$ of $\Delta$ around zero. Dropping the index $\mu$ we write,
\begin{equation}
  F(\Delta) = F_0 + \frac{1}{2}\Upsilon\Delta^2,\quad |\Delta| < r.
  \label{FtoUps0}
\end{equation}
This is trivial in principle, but some complications arise when this is applied
to simulation data with limited accuracy. The following sections will discuss
this question in some detail.

\subsection{Range of $\Delta$}
\label{RangeofDelta}

Since $\Upsilon$ is defined as a derivative of the free energy, the range of
$\Delta$ used for the fit to \Eq{FtoUps0} should be chosen as small as possible.
To check for the dependence of $\Upsilon$ on the range $r$ we made use of a
twist histogram $P(\Delta)$ for an ordinary 3D XY model, with $L=8$ and $T=2.2$
close to $T_c$. It is then found that there is a strong dependence on $r$ which
to a good approximation is $\Upsilon(r) - \Upsilon(0) \sim -r^2$, due to the
presence of a $\Delta^4$ term in $F(\Delta)$. From $\Upsilon(r)$ for small $r$
an extrapolation to $r=0$ gives $\Upsilon = 0.1389(3)$ in excellent agreement
with the more precise value, $\Upsilon=0.13899(8)$ obtained with \Eq{Ups-std}
from a MC simulation with the Wolff cluster algorithm.

\subsection{Disordered systems: unknown $\Delta^{0}$\label{Sec:bias}}

We have so far only been concerned with models with the known minimizing twist
$\Delta^{0} = 0$. The presence of disorder may however mean that the
minimizing twist becomes different for different disorder realizations and is
not known at the outset, and this turns out to add an unexpected complication to
the analysis.

In the case when $\Delta^{0}$ is unknown the analysis consists of two steps:
(i) take some data from a certain range around the maximum of $P(\Delta)$ (the
minimum of the free energy) (ii) fit the free energy from this data to a second
order polynomial in $\Delta$ to determine $\Upsilon$.  For this second step
\Eq{FtoUps0} has to be changed to
\begin{equation}
  F(\Delta) = F_0 + \frac{1}{2}\Upsilon(\Delta-\Delta^{0})^2,\quad
  |\Delta - \Delta^{0}| < r.
  \label{FtoUps}
\end{equation}
When used on simulation data, where statistical fluctuations are always present,
this method happens to give values of the helicity modulus that are biased
towards too large values.

\begin{figure}[htbp]
  \includegraphics[width=8cm]{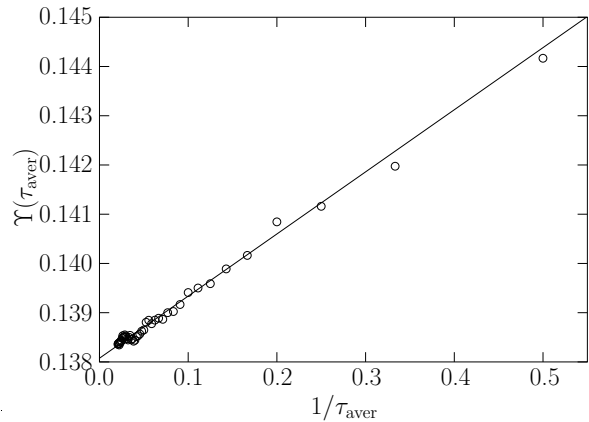}
  \caption{When $\Upsilon$ is determined from \protect\Eq{FtoUps} with
    $\Delta^{0}$ as a free parameter, the obtained $\Upsilon$ becomes biased
    towards too large values. This bias decays with $1/\taver$ where $\taver$ is
    the number of bins used for collecting data, cf.~\Eq{Ptaver}. The data is
    obtained from an isotropic lattice with $L=8$ and $T=2.2$.}
  \label{Upsilon-taver}
\end{figure}

To illustrate this fact we have again made use of twist histograms for the 3D XY
model. Even though the minimizing twist is still zero we now take $\Delta^{0}$
to be a free variable in the analysis.  For the complete run which consists of
about 7000 bins there is no discernible effect of the randomness, but by
constructing twist histograms from $\taver$ (say 2--40) consecutive bins the
effect becomes significant and may be systematically examined. The bin size is
$2^{18} = 262144$ sweeps across the system.  \Figure{Upsilon-taver} shows
$\Upsilon(\taver)$ versus $1/\taver$. The values of these run-length dependent
values $\Upsilon(\taver)$ are based on close to 7000 different twist histograms
constructed from $\taver$ consecutive bins:
\begin{equation}
  \overline{P}^{(i)}(\Delta;\taver) =
  \frac{1}{\taver}\sum_{\tau=1}^{\taver} P(\Delta; i+\tau).
  \label{Ptaver}
\end{equation}
The message from this figure is that there is a bias in determinations of
$\Upsilon$ that are based on too short runs. It is also clear that there is a
$1/\taver$-dependence and the data may be extrapolated to
$\Upsilon(\taver\rightarrow\infty) \approx 0.1381$. Since this data is obtained
with a finite range, $r=0.0625$, this value should be compared to, and agrees
well with, the corresponding value obtained in Sec.~\ref{RangeofDelta}.

A clue to the origin of this bias is given by examining $\Delta^{0}$ which is
the location of the minimum of the free energy. Since the ground state for the
pure 3D XY model is a state with zero twist, $\Delta^{0}=0$, the deviations
from zero are due to the statistical fluctuations in the twist histograms. We
find $\langle (\Delta^{0})^2\rangle \sim 1/\taver$ which is the same as the
behavior of an average $\overline{x}$ of $N$ independent values $x_i$ from a
distribution with zero mean: $\langle \overline{x}^2\rangle \sim 1/N$.

The key observation is now that $\Upsilon$ as a measure of the curvature of the
free energy is inversely related to the width of the distribution. For $N$
samples $x_i$ the width of the distribution is characterized by the variance and
from elementary texts in statistics it is well known that
$\sigma_\mathrm{naive}^2 = \frac{1}{N} \sum_i (x_i - \overline{x})^2$ gives a
biased estimate which may be corrected by
\begin{displaymath}
  \sigma^2 = \frac{N}{N-1} \sigma_\mathrm{naive}^2.
\end{displaymath}
The well known reason for this correction is the use of $\overline{x}$ instead
of the true average of the distribution. In the analysis of the histogram data
the location of $\Delta^{0}$ is a similar source of error and it is natural to
expect the same kind of effect in the analysis of the histogram data. With
$\Upsilon(\taver)$ and $\Upsilon(\infty)$ inversely related to
$\sigma_\mathrm{naive}^2$ and $\sigma^2$, respectively, and the number of
independent samples given by $N = \taver/b$, we obtain
\begin{equation}
  \Upsilon(\taver) = \frac{1}{1-b/\taver} \Upsilon(\infty).
  \label{Upsilon-bias}
\end{equation}
Here $b$ is a constant with dimension of time. The expression above may also be
written
\begin{equation}
  \frac{1}{\Upsilon(\taver)} = \frac{1}{\Upsilon(\infty)} - b'/\taver,
  \label{Upsilon-inv-bias}
\end{equation}
and for small values of $b/\taver$ \Eq{Upsilon-bias} becomes
\begin{equation}
  \Upsilon(\taver) = \Upsilon(\infty) + b''/\taver,
  \label{Upsilon-linear-bias}
\end{equation}
which explains the rectilinear behavior of $\Upsilon$ in \Fig{Upsilon-taver}. To
determine the unbiased quantity $\Upsilon(\infty)$ we need to obtain
$\Upsilon(\taver)$ for a few values of $\taver$ and fit that data to one of the
equations above.

\subsection{Twist fluctuations in several directions}

We have now discussed the use of twist fluctuations in a single direction and
ordinary PBC in the two other. The simulations of \VG\ were however done in a
somewhat different way with twist fluctuations in all three directions. For this
discussion we introduce the generalization $P^{(3)}(\Delta_x, \Delta_y,
\Delta_z)$. With the phase angles of the XY model discretized to 256 different
values, the computer memory needed to store such histograms rapidly becomes
enormous. The collected histograms were therefore instead
\begin{displaymath}
  P_x(\Delta_x) = \sum_{\Delta_y} \sum_{\Delta_z} P^{(3)}(\Delta_x,\Delta_y, \Delta_z),
\end{displaymath}
and the analogous $P_y(\Delta_y)$ and $P_z(\Delta_z)$.  For the quantity defined
earlier with twist fluctuations in one dimension only we write,
\begin{displaymath}
  P^{(1)}(\Delta_x) \equiv P^{(3)}(\Delta_x,0,0).
\end{displaymath}

\begin{figure}[htbp]
  \includegraphics[width=8cm]{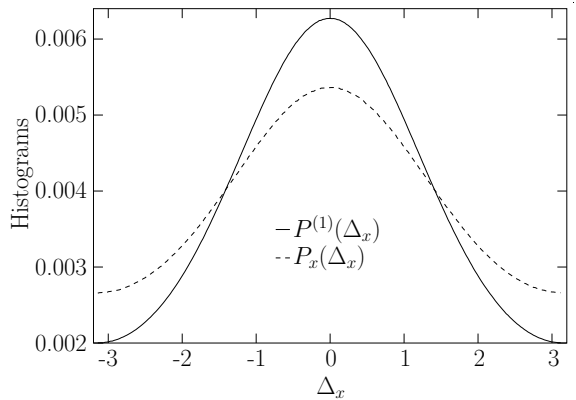}
  \caption{Comparison of the twist histograms from simulations of the pure 3D XY
    model with twist fluctuations in one and three directions. The solid line is
    the distribution of $\Delta_x$ in simulations with $\Delta_y=\Delta_z=0$.
    The dashed line is the same quantity obtained with fluctuations in all three
    directions. The data is obtained from a cubic isotropic lattice with $L=8$
    and $T=2.2$.}
  \label{cmp-tw-twx}
\end{figure}

\begin{figure}[htbp]
  \includegraphics[width=8cm]{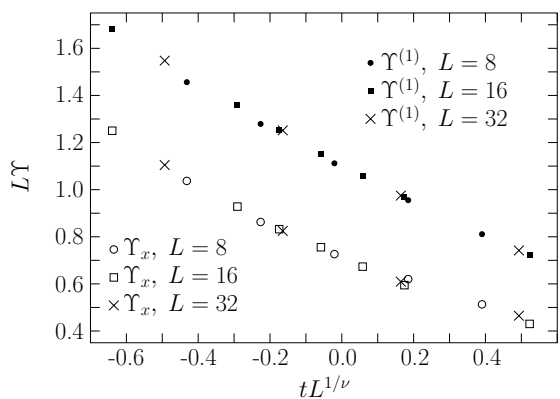}
  \caption{Scaling collapse of the helicity modulii obtained in the pure 3D XY
    model. The upper symbols (solid) are the proper helicity modulus,
    $\Upsilon^{(1)}$ obtained from \Eq{Ups-std} and the low symbols are for
    $\Upsilon_x$ obtained from $P_x(\Delta_x)$. The good collapse of the latter
    quantity confirms the expectation that it equally well may be used for
    examining the critical properties.}
  \label{Upsilon-twist3}
\end{figure}

\Figure{cmp-tw-twx} shows $P^{(1)}(\Delta_x)$ together with $P_x(\Delta_x)$.
These two curves are very different and it becomes clear that the ``helicity
modulus'' determined from $P_x$ is not the same as the proper helicity modulus
from the fluctuation formula or from $P^{(1)}$. However, from the universality
hypothesis one would expect scaling of all kinds of quantities based on the free
energy, and the similar behavior of $\Upsilon_x$ and $\Upsilon^{(1)}$ in
\Fig{Upsilon-twist3} suggests that that actually is the case. Here we use the
standard scaling assumption,
\begin{equation}
  \label{eq:scale}
  L\Upsilon = f_\Upsilon(t L^{1/\nu}),
\end{equation}
with the reduced temperature $t = (T/T_c-1)$.

\subsection{Use in vortex glass simulations}

As discussed above there are some complications in the determination of the
helicity modulus from twist histograms. However, when our interest is only to
determine the critical properties of a model, two of the above discussed
complications may be disregarded. If the scaling hypothesis is phrased such that
the properties of the free energy is a function only of the combination
$tL^{1/\nu}$ it is clear that the precise method to examine these properties is
not important as long as it is the same for all system sizes.  Among other
things this means that the choice of the range $r$ of $P_\mu(\Delta_\mu)$ used
for determining the helicity modulus is immaterial.  Similarly, the difference
between the proper helicity modulus and the quantity obtained from $P_\mu$ need
not bother us either. The crucial point that has to be taken care of is the
elimination of the bias of Sec.~\ref{Sec:bias} since this bias (as shown in
\Fig{Uperp-inv-taver}) is different for different $L$.

When considering disordered systems there is one more point that should be taken
under consideration. The parameter $b$ in \Eq{Upsilon-bias} has the dimension of
time and may be interpreted as the time between two independent measurements.
In a disordered system one expects the characteristic time to be different for
different disorder realizations and one would need an average of a number of
functions with different time constants. However, since the correction is linear
in $b$, c.f.\ \Eq{Upsilon-linear-bias}, such an average has the same functional
form, but now with $b$ as an \emph{average} characteristic time.

\section{The vortex glass model}

The model we simulate is given by the
Hamiltonian\cite{Olsson_Teitel:xy3fp,Olsson:vg-lett}
\begin{equation}
  {\cal H} = -\sum_{{\rm bonds}\,i\mu}J_{i\mu}\cos
  (\theta_i-\theta_{i+\hat\mu}-A_{i\mu} + \delta_\mu^{(\mathbf{r}_i\cdot \hat{\mu})}),
  \label{Hamiltonian}
\end{equation}
where $\theta_i$ is the phase of the superconducting wave function at site $i$
with position $\mathbf{r}_i$ of a periodic $L_x\times L_y\times L_z$ lattice,
and the sum is over all bonds in directions $\mu = x$, $y$, $z$.  The size in
the $x$ and $y$ directions are the same; $L_x = L_y = L$. An applied magnetic
field in the $z$ direction is obtained through the quenched vector potential
with the choice $A_{ix} = 2\pi f y_i$, and $A_{iy} = A_{iz} = 0$. The
simulations are performed with fluctuating twist boundary
conditions\cite{Olsson:self-cons.long} which in the duality relation corresponds
to a vortex line model with periodic boundary conditions.  We make use of
$L_\mu$ twist variables $\delta_\mu^{(\mathbf{r}_i\cdot \hat{\mu})}$ in each
direction and the total twists in the respective directions are $\Delta_\mu
=\sum_{j=1}^{L_\mu}\delta_\mu^{(j)}$.\cite{why-this-delta} These variables are
updated with the usual Metropolis method. We have run simulations for
four different sets of parameters, summarized in \Tabsim. The disorder is put in
the coupling constants which are chosen as
\begin{eqnarray*}
  J_{i\mu} &= &\Jperp(1+\varepsilon_{i\mu}), \quad \mu=x,y, \\
  J_{iz} &= &\Jpara(1+\varepsilon_{i\mu}), \quad \mbox{only sets C and D}.
\end{eqnarray*}
For set A the $\varepsilon_{i\mu}$ are independent variables from a Gaussian
distribution with $\langle\varepsilon_{i\mu}\rangle = 0$ and $p =
\sqrt{\langle\varepsilon_{i\mu}\rangle^2} = 0.40$. For sets B through D
$\varepsilon_{i\mu}$ were instead from a uniform distribution between $-1$ and
$1$. Another difference (as indicated in the Table) is that the disorder for
sets A and B is only put on the couplings in the $x$ and $y$ directions whereas
the sets C and D are also disordered along $z$. The reason for this choice is to
facilitate a direct comparison with the simulations in \onlineRef{Kawamura:03}.

\begin{table}[htbp]
  \begin{tabular}{|c|cclc|}
     Set & $f$ & $\Jpara/J_\perp$ & disorder & directions \\
    \hline
    A & $1/5$ & $1/40$ & Gaussian & $x$, $y$ \\
    B & $1/5$ & $1/10$ & rectangular & $x$, $y$ \\
    \hline
    C & $1/4$ & $1/10$ & rectangular & $x$, $y$, $z$ \\
    D & $1/4$ & $1$    & rectangular & $x$, $y$, $z$
  \end{tabular}
  \caption{Four different parameter sets have been simulated. Information about
    the runs are given in Tab.~\ref{tab:runs}.}
  \label{tab:sets}
\end{table}

\section{Simulation methods}

\subsection{The exchange steps}

The exchange MC method---also called parallel tempering---is an elegant method
that makes it possible to calculate the correct statistical averages in
disordered systems where the usual MC methods would only be stuck in a local
minimum. The idea is to simulate many different configuration in parallel and,
beside the ordinary Metropolis MC steps, let the configurations perform a kind
of constrained random walk in temperature space. These occasional changes in
temperature means that the configurations sometimes are at higher temperatures
where the energy barriers between various local minima are low and easily may be
overcome.

Our simulations were done with $N_T$ temperatures, $T_0$ through $T_{N_T-1}$,
chosen according to
\begin{equation}
  T_m = \Tmin \left(\frac{\Tmax}{\Tmin}\right)^{m/N_T},\quad
  m=0,\ldots,N_T-1.
  \label{Tm}
\end{equation}
The values of $N_T$, $\Tmin$, and $\Tmax$ as well as the number of disorder
realizations and the length of the runs are detailed in Tab.\ \ref{tab:runs}

\begin{table}[htb]
  \begin{tabular}{|c|r|c|r|c|c|r|r|}
   Data set & $L$& $N_d$ & $N_T$ & $\Tmin$ & $\Tmin$ &
   $\teq$ & $\tmax$\\
   \hline
   A
   & $10$ & 600 & 12 & 0.09  & 0.24 &  1 & 16 \\
   $L_z/L=3/5$                      
   & $15$ & 600 & 24 & 0.09  & 0.24 &  4 & 16 \\
   & $20$ & 600 & 36 & 0.09  & 0.24 & 11 & 32 \\
   & $25$ & 200 & 36 & 0.115 & 0.24 & 17 & 48 \\
   \hline                           
   A                                
   & $10$ & 900 & 12 & 0.09  & 0.24 &  2 & 13 \\
   $L_z/L=2/5$                      
   & $15$ & 900 & 24 & 0.09  & 0.24 &  5 & 17 \\
   & $20$ & 460 & 36 & 0.09  & 0.24 & 12 & 33 \\
   \hline                           
   B                                
   & $10$ & 500 & 12 & 0.18  & 0.40 &  3 & 13 \\
   & $15$ & 500 & 24 & 0.18  & 0.40 &  5 & 21 \\
   & $20$ & 300 & 36 & 0.18  & 0.40 &  7 & 21 \\
   \hline                           
   C                                
   &  $8$ & 400 &  8 & 0.16  & 0.38 &  1 & 12 \\
   & $12$ & 700 & 16 & 0.16  & 0.38 &  2 & 15 \\
   & $16$ & 400 & 24 & 0.16  & 0.38 &  3 & 17 \\
   \hline                           
   D                                
   &  $8$ & 400 &  8 & 0.55  & 1.10 &  1 & 13 \\
   & $12$ & 600 & 16 & 0.55  & 1.10 &  2 & 15 \\
   & $16$ & 600 & 24 & 0.55  & 1.10 &  3 & 17 \\
   & $20$ & 400 & 32 & 0.55  & 1.10 &  4 & 13 \\
   \hline
  \end{tabular}
  \caption{Parameters describing the simulations. For systems of size
   $L\times L\times L_z$ we simulated $N_d$ disorder configurations
   with $N_T$ temperatures in the range $\Tmin \leq T < \Tmax$, cf.\
   \Eq{Tm}. Of the bins corresponding to $2^{18}=262144$ sweeps, $\teq$ are first
   discarded and the remaining $\tmax-\teq$ are used for calculating averages.} 
  \label{tab:runs}
\end{table}

\subsection{Check for equilibration}

\begin{figure}[htbp]
  \includegraphics[width=8cm]{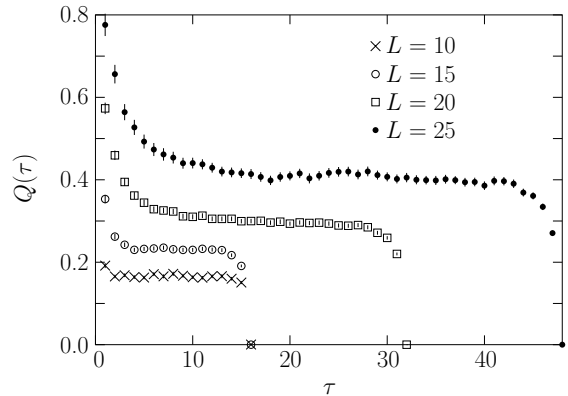}
  \caption{The quantity $Q_\mu(\tau)$ is a disorder averaged measure of
    the difference between the histograms $P_\mu(\Delta_\mu,\tau)$ and
    $P_\mu(\Delta_\mu, \tmax)$. The initial decrease down to a constant level
    shows the thermalization of the collection of $N_T$ configurations. The
    decrease as $\tau\rightarrow\tmax$ is there because of correlations between
    $P_\mu(\Delta_\mu, \tau)$ for consecutive $\tau$. This data is for set A
    with $L_z/L=3/5$; the shown quantity is $Q = \frac{1}{2}(Q_x + Q_y)$.}
  \label{fig:Q-t}
\end{figure}

In spite of its beauty the exchange MC method does not alleviate the need for
thermalizing the system and it is therefore necessary to in some way monitor the
approach to equilibrium. Since our main quantities from the simulations are the
histograms $P_\mu(\Delta_\mu, \tau)$ we use these quantities in the analysis of
the approach to equilibrium. The idea is to quantify the similarity of each
histogram $P_\mu(\Delta_\mu, \tau)$ to the last histogram $P_\mu(\Delta_\mu,
\tmax)$ which is assumed to be typical of a thermalized system. The disorder
averaged histogram difference is then defined as
\begin{equation}
  Q_\mu(\tau) = \av{\sum_{\Delta_\mu}
    \left|P_\mu(\Delta_\mu, \tau) - P_\mu(\Delta_\mu, \tmax)\right|}
  \label{Qtau}
\end{equation}
The notation $\av{\ldots}$ denotes the disorder averaging.  $Q = \frac{1}{2}(Q_x
+ Q_y)$ is shown in \Fig{fig:Q-t} for $T=0.125$ (close to $T_c$) and our four
system sizes.  The decrease of $Q$ at small $\tau$ down to constant levels is an
effect of the thermalization and one can read off the number of bins needed for
thermalization ($\teq$ in Tab.~\ref{tab:runs}) from \Fig{fig:Q-t}. The decrease
of $Q$ as $\tau\rightarrow\tmax$ is due to similarities between
$P_\mu(\Delta_\mu, \tau)$ and $P_\mu(\Delta_\mu, \tmax)$ that are present
because of the slow dynamics of the MC simulations. The equilibration time
$\teq$ is chosen from the time needed to reach the constant value of $Q(\tau)$
whereas $\tmax$, the total length of the simulations, was chosen to get enough
data for the extrapolation shown in \Fig{Uperp-inv-taver}.

The constant level of $Q(\tau)$ is, especially for the larger systems,
surprisingly large. To interpret the data correctly it should however be kept in
mind that the figure shows the difference between two histograms that both
deviate from the true histogram of a hypothetical run of infinite length; the
difference between a single histogram and the true one would give values that
are roughly a factor of two smaller. The values nevertheless signal large
fluctuations and conveys a message about a complicated phase space with many
different local minimas.

\subsection{Efficiency of the exchange steps}

A common way to monitor the efficiency of the exchange MC steps is to measure
the exchange acceptance. This is however only a measure of the local
mobility of the configurations and doesn't answer the more relevant question
about the efficiency of the algorithm to move configurations across a larger
temperature range. To keep track of all the exchange steps would mean producing
an enormous amount of data and is therefore usually not desirable. A simple
method has therefore been devised that gives the most relevant information with
very little overhead. The idea is to, for each configuration, keep track of the
time since the visit at each given temperature. To that end each configuration
is accompanied by a vector of integers, $v_m$, with information about how long
it was since the temperature $T_m$ was last visited by that very configuration.

\begin{figure}[htbp]
  \includegraphics[width=8cm]{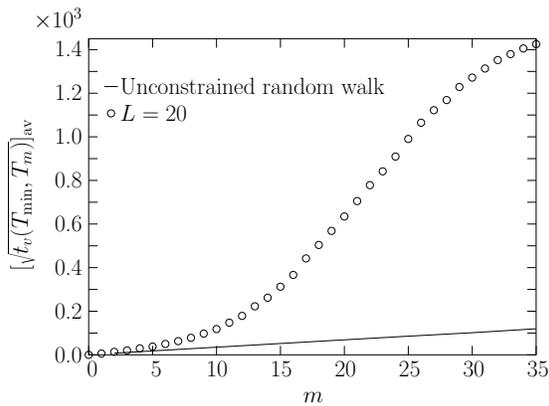}
  \caption{The quantity $t_v(T_\mathrm{min}, T_m)$ is the time, measured in
    number of sweeps, for a configuration to travel from temperature $T_m$ down
    to $T_\mathrm{min}\equiv T_0$. The open symbols show the disorder average of
    the square root of $t_v$ based on 600 disorder configurations. The solid
    line is an estimate of the same quantity from the exchange acceptance. From
    the large difference it is clear that conclusions about the efficiency of
    the exchange steps to make the configurations travel across large
    temperature regions should not be drawn based on the acceptance ratio alone.
    We plot $\sqrt{t_v}$ since this quantity is proportional to the distance for
    a simple random walk.  }
  \label{fig:tv20}
\end{figure}

One way to use that information is to examine the vectors $v_m$ for all
configurations that were at the lowest temperature at the end of the run. A
measure of the disorder averaged time since the last visit at temperature $T_m$
is shown in \Fig{fig:tv20}. The same figure also shows the results from a simple
simulation of an unconstrained random walk with the same properties and
acceptance probability as in the exchange MC.  As seen in the figure the
difference is an order of magnitude, which indicates that conclusions about the
efficiency of the exchange steps cannot be safely determined from the acceptance
ratio alone. The reason for the long times needed for a configuration to travel
from the highest to the lowest temperature is presumably that most
high-temperature states are far away from the phase space regions typical of the
lowest temperatures, which means that the configuration will usually have to
undergo many thorough and time-consuming reorganizations before it can reach an
energy compatible with the lowest temperatures.

\subsection{Eliminating the bias}

\begin{figure}[htbp]
  \includegraphics[width=8cm]{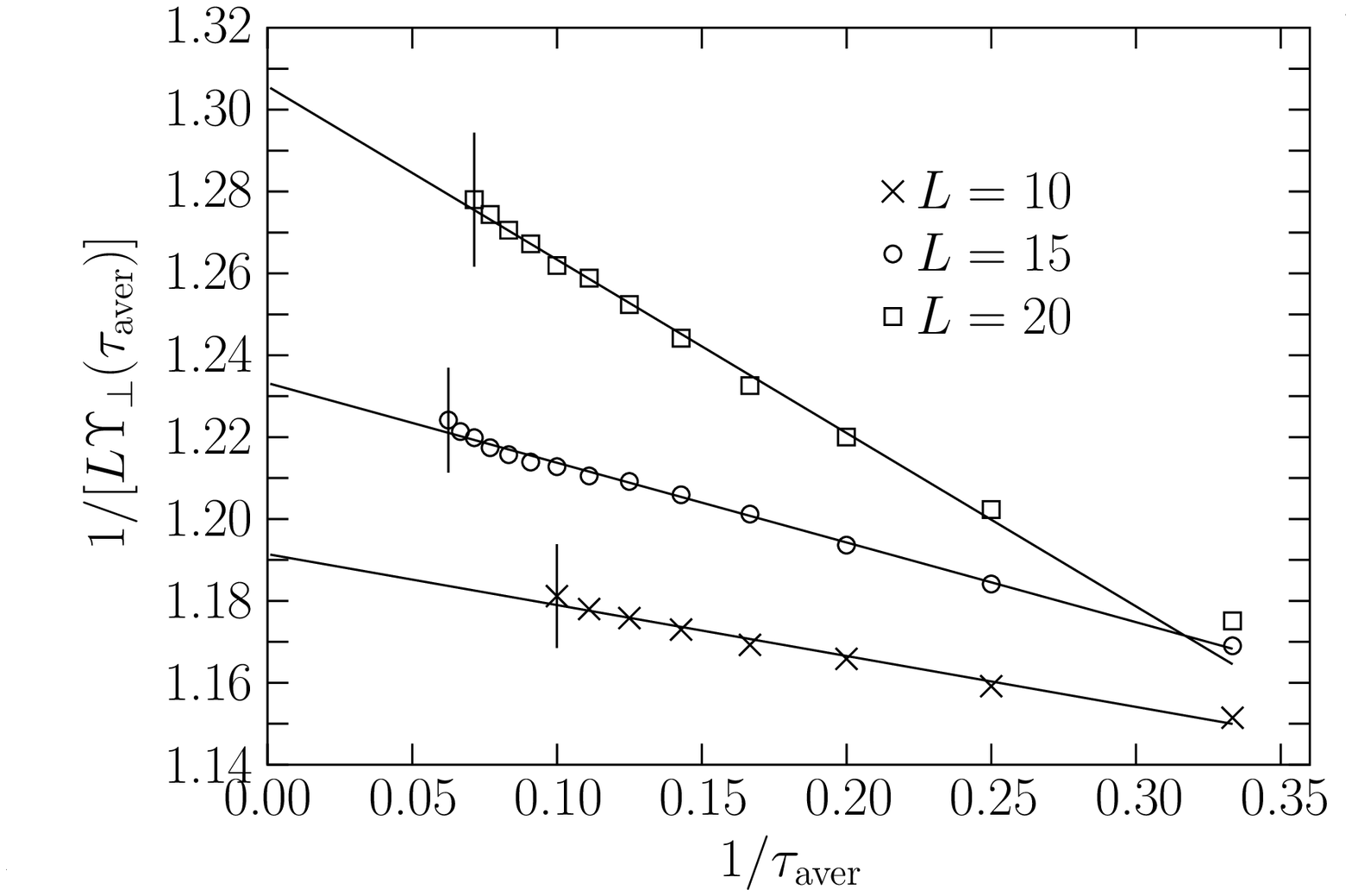}
  \caption{The elimination of the bias in $\Uperp(\taver)$ is done by
    extrapolating to $\taver\rightarrow\infty$ with \Eq{Upsilon-inv-bias}. The
    present data is from set B, $T=0.2685$ closely above $T_c$. }
  \label{Uperp-inv-taver}
\end{figure}

For the following discussion we introduce a notation for the disorder averaged
helicity modulii in the transverse and the parallel directions, respectively,
\begin{subequations}
  \label{DisAver}
\begin{eqnarray}
  \Uperp & = & \frac{1}{2}\av{\Upsilon_x + \Upsilon_y}, \\
  \Upara & = & \av{\Upsilon_z}.
\end{eqnarray}
\end{subequations}
The procedure used to determine the disorder averaged helicity modulus consists
of three steps: (i) Determine $\Upsilon_\mu(\taver)$ for each disorder
configuration and several values of $\taver$ by fitting histogram
$\overline{P}_\mu(\Delta_\mu;\taver)$ based on $\taver$ consecutive bins,
$P_\mu(\Delta_\mu,\tau)$ to \Eq{FtoUps0}. (ii) Calculate the disorder averaged
quantities $\Uperp(\taver)$ and $\Upara(\taver)$, cf.\ \Eqs{DisAver}.  (iii) Fit
this data to \Eq{Upsilon-inv-bias} to obtain the unbiased estimates $\Uperp
\equiv \Uperp(\infty)$ and $\Upara \equiv \Upara(\infty)$.  The last step is
illustrated in \Fig{Uperp-inv-taver}.  The error bars are the statistical errors
associated with the disorder average for each size. It seems that the errors
associated with the extrapolation to zero $1/\taver$ are smaller than the errors
due to the limited number of disorder realizations.

\section{Results}
\label{sec:results}

In this section we report the results from the analysis described above, with a
number of different sets of parameters, cf.~\Tabsim.  The purpose is to check
that the proposed behavior is a generic feature and is not limited to the
parameters of \VG, but as we have been doing simulations with several different
sets of parameters it has not been possible to achieve very high precision in
the estimates of the critical exponents. The emphasis is therefore rather on
checking for scaling that is consistent with 3D gauge glass universality.
Generally speaking that picture is confirmed, but the new simulations also give
information about failure of finite size scaling for certain sizes and
parameters.

\subsection{High anisotropy, $\Jpara/J_\perp = 1/40$}

The results in \VG\ were obtained with a rather high anisotropy, $\Jpara/J_\perp
= 1/40$ and the aspect ratio $L_z/L=3/5$. We have now also performed simulations
with a smaller aspect ratio, $L_z/L = 2/5$, which is a good consistency test
since not only the critical exponents but also the critical temperature should
be independent of the aspect ratio. We have also performed additional
simulations with several other aspect ratios to determine the anisotropy
exponent.

\begin{figure}[htbp]
  \includegraphics[width=8cm]{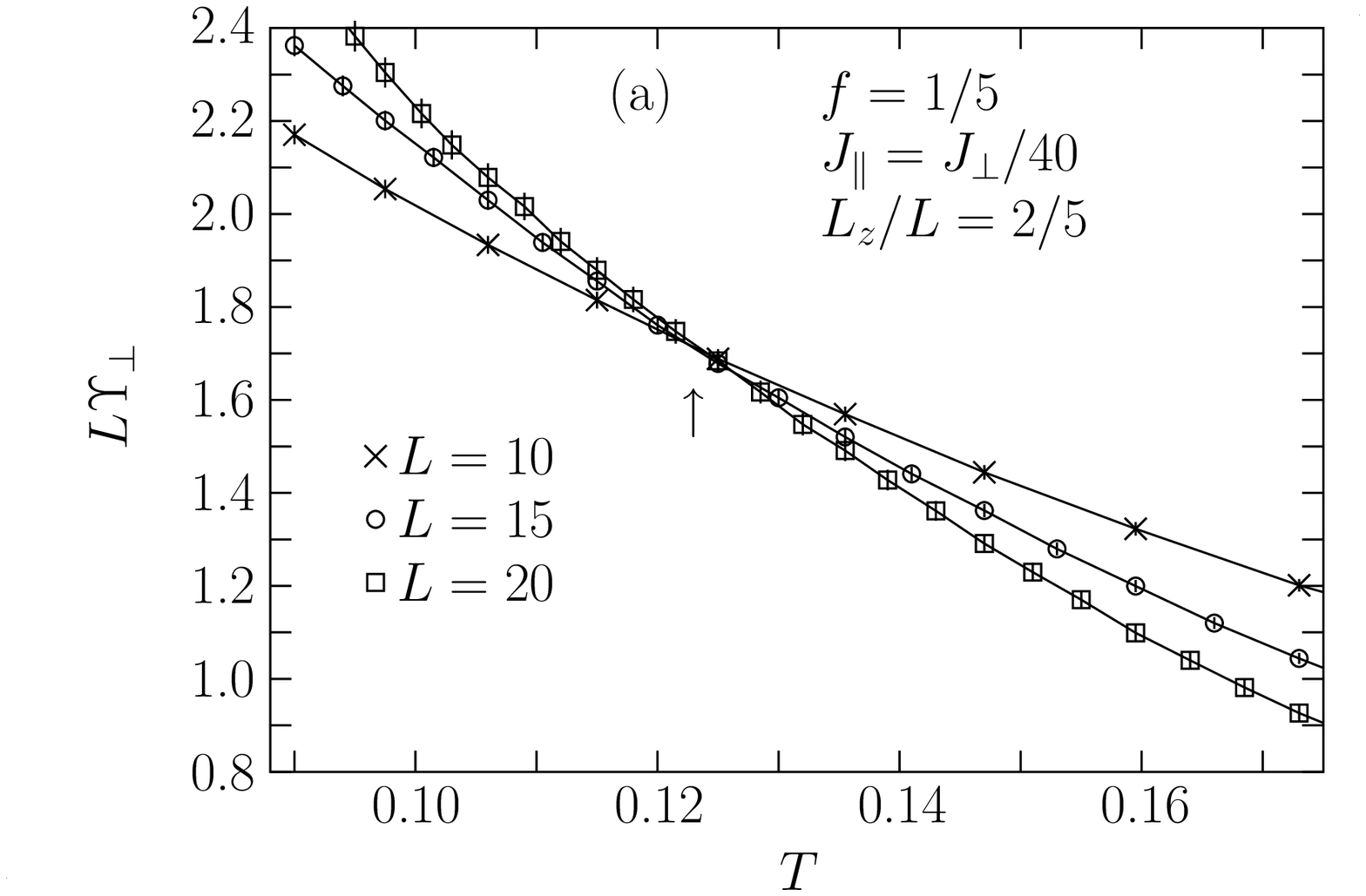}
  \includegraphics[width=8cm]{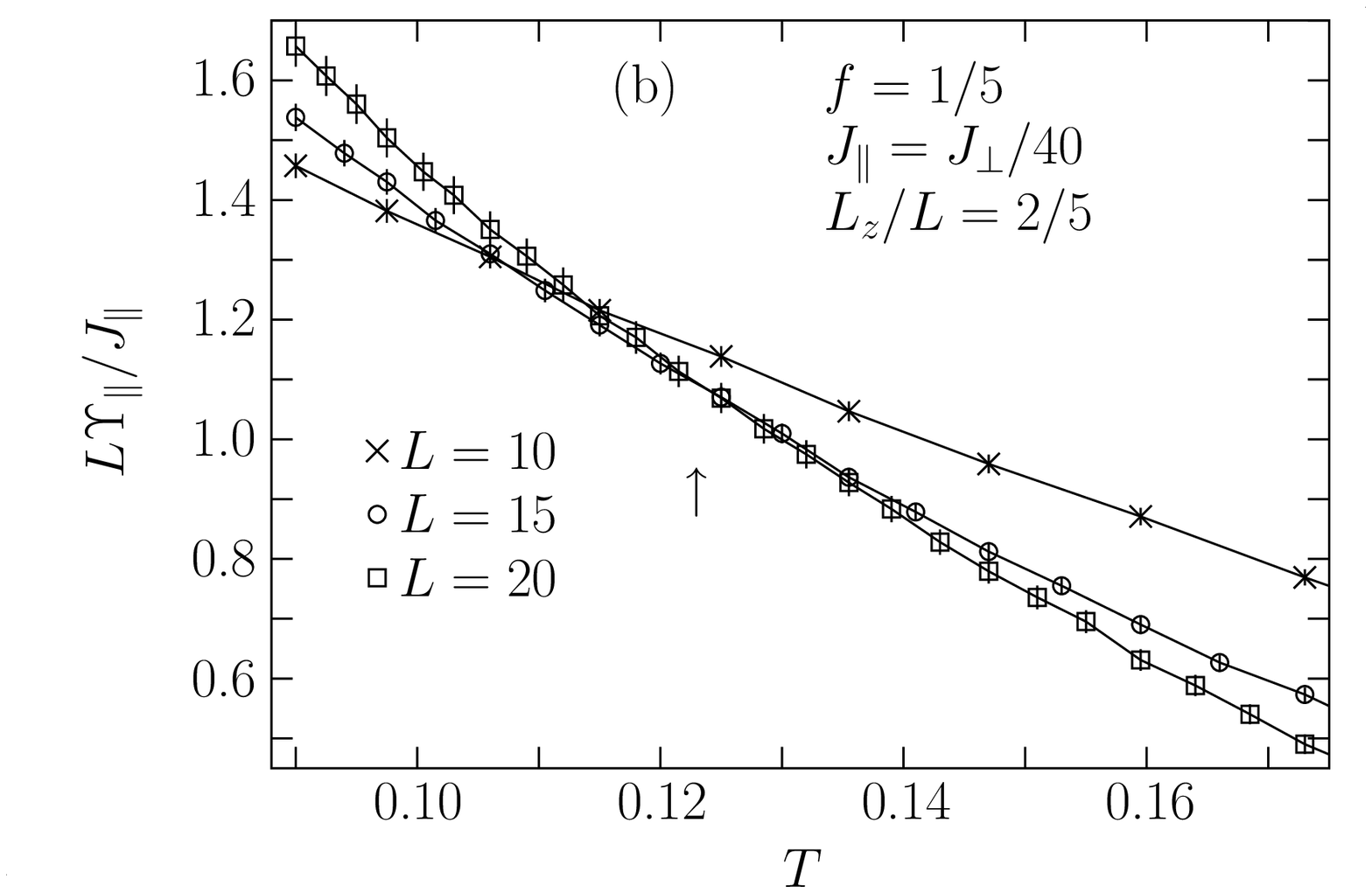}
  \caption{Helicity modulii from simulations with aspect ratio
    $L_z/L=2/5$. The data for $L\Uperp$ in panel (a) all cross at $T_c\approx
    0.123$ (shown by the arrow) in agreement with \VG. Panel (b) for $L\Upara$
    on the other hand shows the expected crossing only for the two largest
    sizes. The deviation from the scaling behavior is another example of the
    well-known fact that finite size scaling often fails for very small system
    sizes.}
  \label{fig:10x04-Ups}
\end{figure}

\subsubsection{Varying the aspect ratio}

\Figure{fig:10x04-Ups} shows the helicity modulii for the same parameters as in
\VG\ but with the aspect ratio $L_z/L=2/5$.  We find a nice crossing for
$L\Uperp$ at the expected value $T_c=0.123$.\cite{Olsson:vg-lett} The results
for the perpendicular quantity $L\Upara$ also agree with this behavior for the
two larger sizes but the data for the smallest system, $10\times10\times4$, is
significantly off. This is in line with the general expectation that the scaling
only should work for rather large system sizes.  However, somewhat unexpectedly,
the scaling in $L\Uperp$ prevails even though it fails in the direction parallel
to the applied field.

\begin{figure}[htbp]
  \includegraphics[width=8cm]{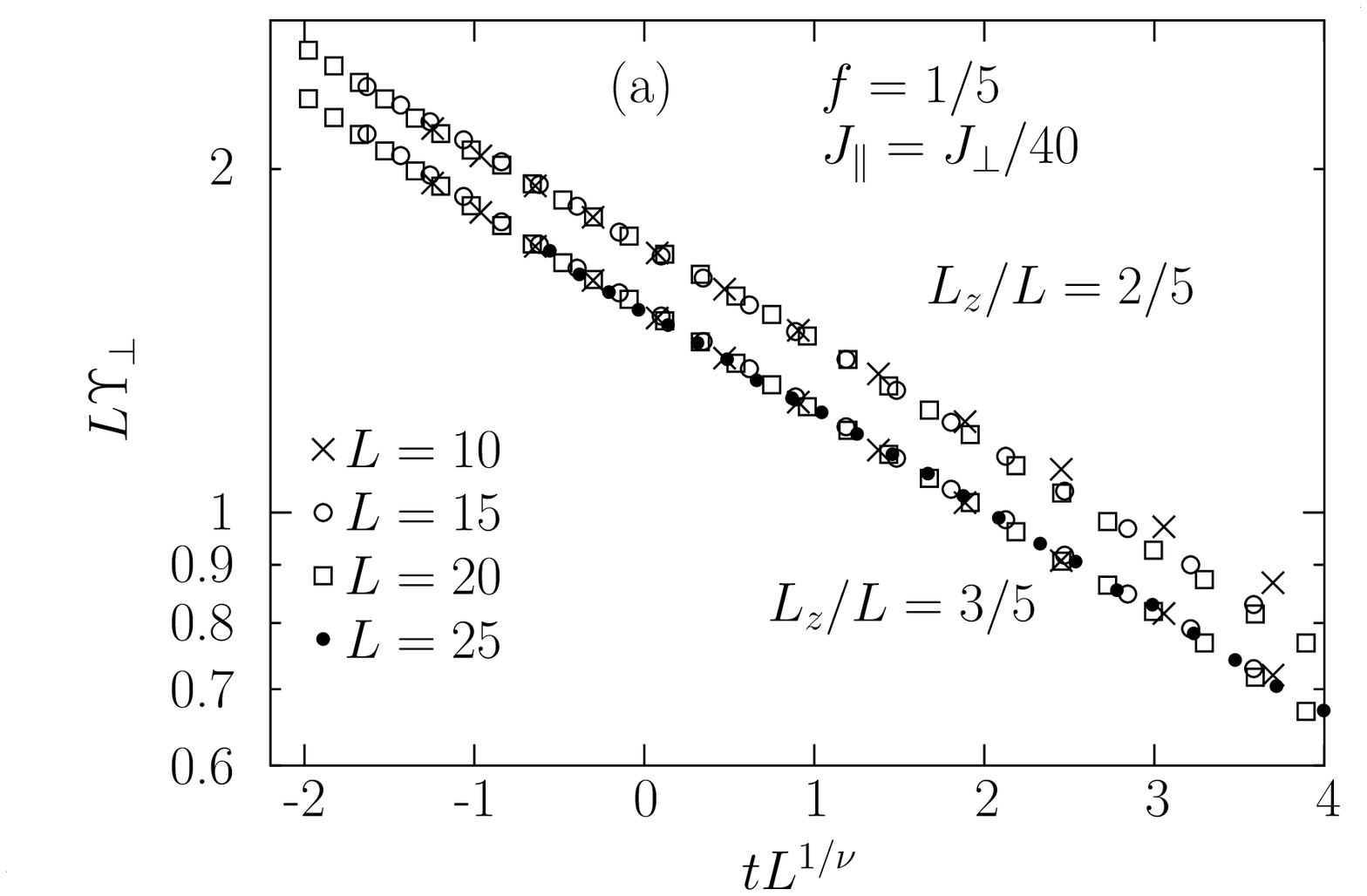}
  \includegraphics[width=8cm]{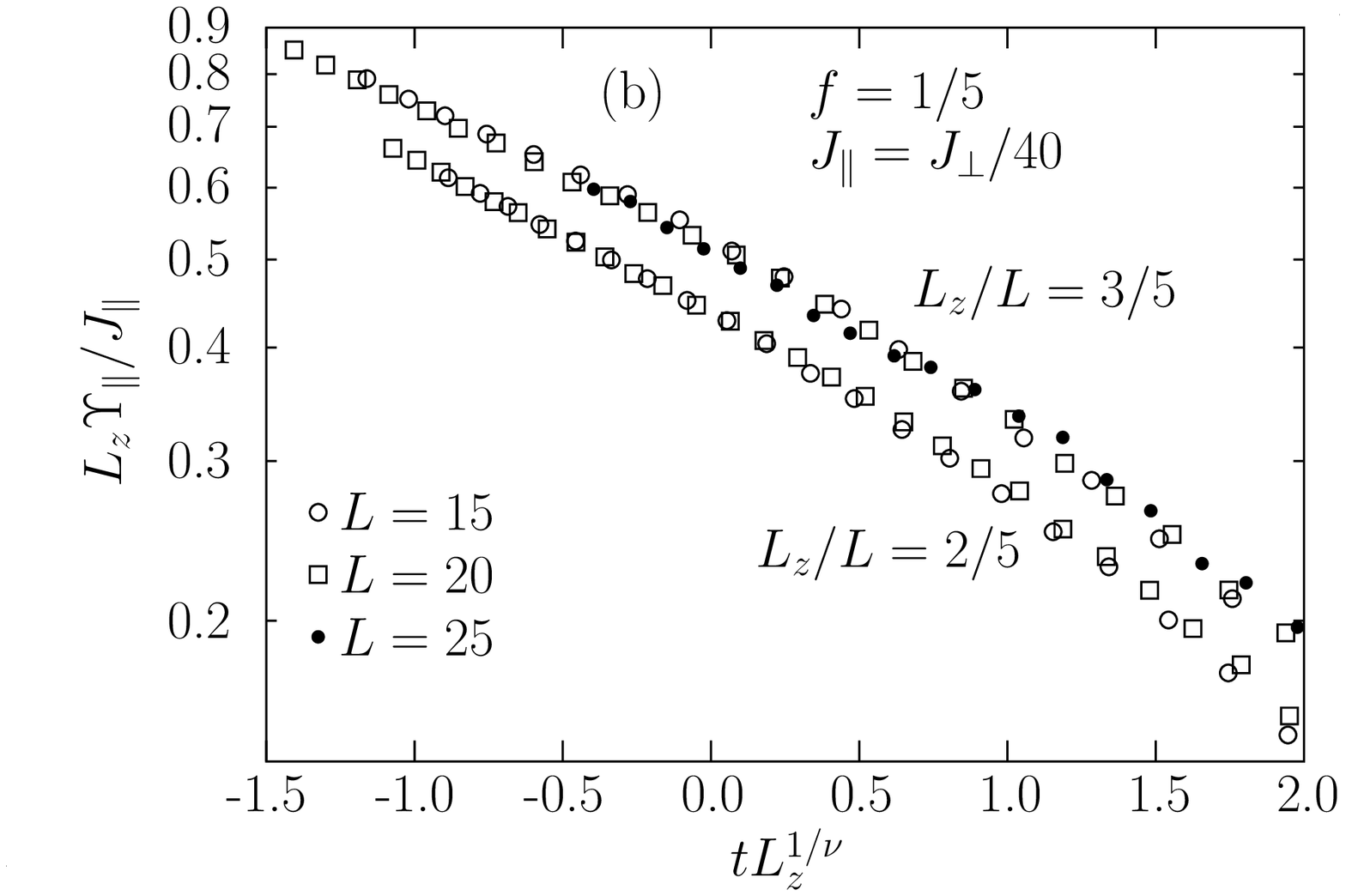}
  \caption{Collapse of $L\Uperp$ and $L_z\Upara$ with $\nu=1.5$ and $T_c=0.123$
    from \VG\ for two different aspect ratios, $L_z/L = 2/5$ and $3/5$. The
    reduced temperature is $t=(T/T_c - 1)$.  For $\Upara$ the data for the
    smallest size, $L=10$, has been omitted since it appears to be too small to
    scale, cf.\ panel (b) in \Fig{fig:10x04-Ups}.}
  \label{fig:10x04-Uperp-aspect}
\end{figure}

Scaling collapses for the two aspect ratios $L_z/L=2/5$ and $3/5$ are shown in
\Fig{fig:10x04-Uperp-aspect}.  When discarding $\Upara$ for $10\times10\times4$
the collapses with $T_c=0.123$ and $\nu=1.5$ (from \VG) are excellent for both
quantities.  Note the similar shapes of the scaling functions for the two aspect
ratios. For panel (b) this requires the use of $L_z$ instead of $L$ on the $x$
axis.  Also note that the dependency on the aspect ratio is the opposite for
$L_z\Upara$ compared to $L\Uperp$.

\subsubsection{The anisotropy exponent}

The above data is consistent with isotropic scaling, the anisotropy exponent
$\zeta=1$, but to estimate the error bars we need a determination of the
exponent.  The idea behind finite size scaling is that certain quantities only
should depend on the fraction $\xi/L$, and to generalize this concept to
anisotropic scaling one has to allow for the possibility of two different
correlation lengths, $\xi$ and $\xi_z$, that grow in different ways as $T_c$ is
approached, $\xi_z \sim \xi^\zeta$. To do finite size scaling one needs sizes
such that $\xi/L \propto \xi_z/L_z$ and with the above relation between $\xi$
and $\xi_z$ we need to determine the behavior of systems with $L_z\propto
L^\zeta$. For general values of $\zeta$ this gives non-integral $L_z$ and the
common practice is to obtain the appropriate data through interpolation of data
for neighboring $L_z$-values. To make that possible we have thus simulated with
several different $L_z$: for $L=10$ we have used $L_z = 5$, 6, and 7, and for
$L=15$ simulations have been done with $L_z = 8$, 9, and 10.

\begin{figure}[htbp]
  \includegraphics[width=8cm]{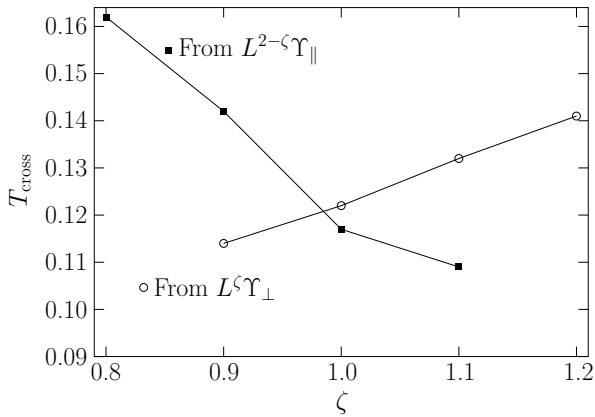}
  \caption{The figure shows how the crossing temperature of
    $L^\zeta\Uperp$ and $L^{2-\zeta}\Upara$ with $L=15$ and $L=25$ depend on the
    assumed value of $\zeta$. One set of data is for $(L,L_z)=(25,15)$ and the
    other for $(L,L_z) = (15,9\cdot(15/25)^{\zeta-1})$.  The second set is
    obtained by interpolating the results from simulations with $L_z = 8$, 9,
    and 10. Since a crossing of the data in both directions should occur at
    $T_c$ the correct value of $\zeta$ is obtained at the crossing of these two
    sets of data points. This gives $\zeta=1\pm0.1$, strongly suggestive of
    isotropic scaling.}
  \label{fig:zeta}
\end{figure}

To determine limits on $\zeta$ the most straightforward test would be to repeat
the scaling analysis with different values of $\zeta$ and check how the quality
of the scaling collapse depends on $\zeta$. Because of the statistical errors in
the raw data that is however not a very useful technique. A more sensitive test
is obtained by combining results from analyses of both $\Uperp$ and $\Upara$. To
do that we focus on how the crossing temperatures of $L^\zeta\Uperp$ and
$L^{2-\zeta}\Upara$ depend on $\zeta$. To make the test clean and simple we only
make use of two sizes at the time.  \Fig{fig:zeta} shows the dependency of the
crossing temperatures on $\zeta$ for sizes $L=15$ and 25. The two different
crossing temperatures coincide at $T\approx0.12$ and $\zeta\approx 1$. Note that
the two quantities have the opposite dependency on $\zeta$. This is the key to
this more precise determination of $\zeta$, and together with a rough error
estimate \Fig{fig:zeta} gives $\zeta=1\pm0.1$.

\subsection{Less anisotropic, $\Jpara/\Jperp = 1/10$}

\begin{figure}[htbp]
  \includegraphics[width=7cm]{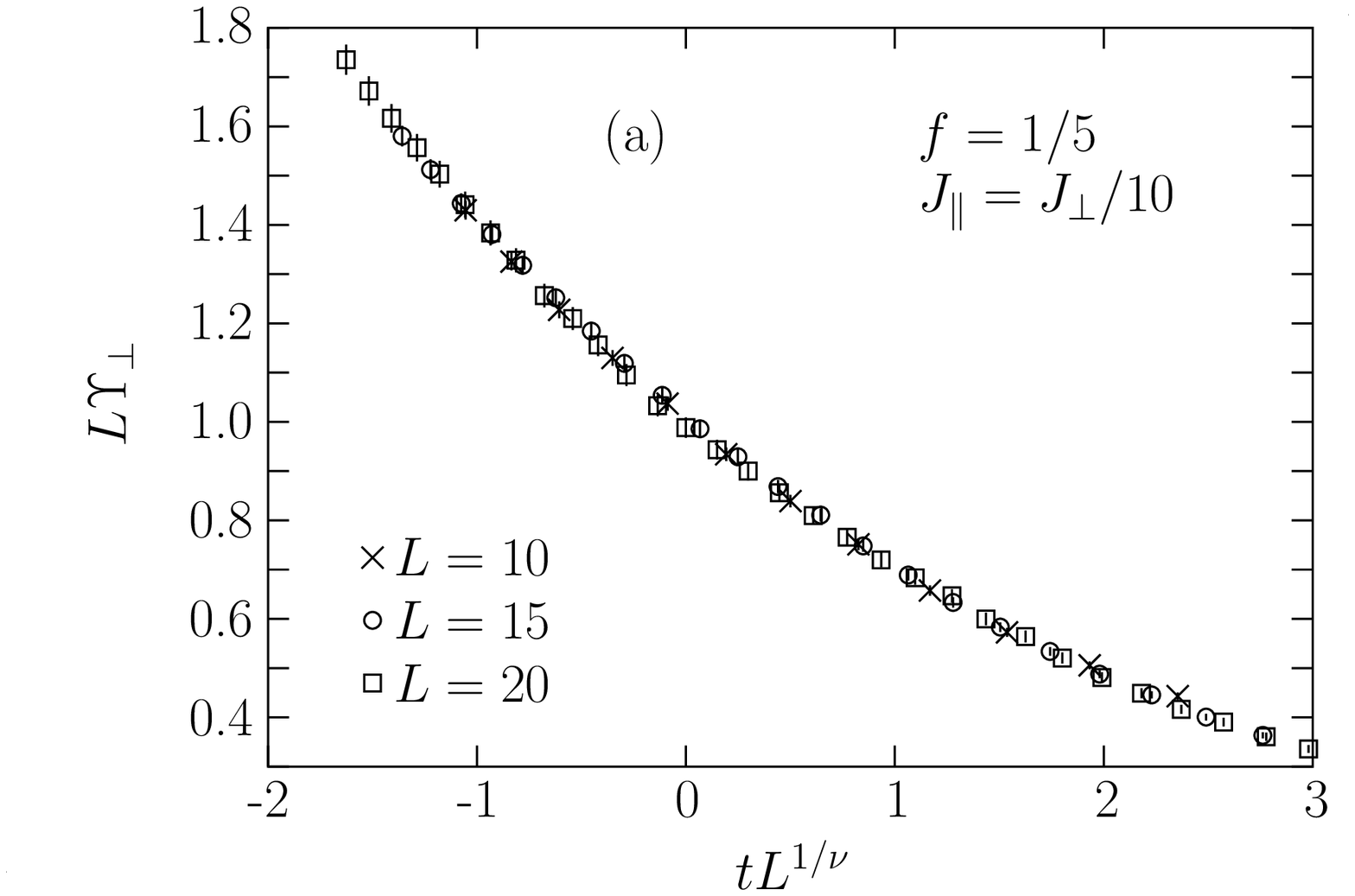}
  \includegraphics[width=7cm]{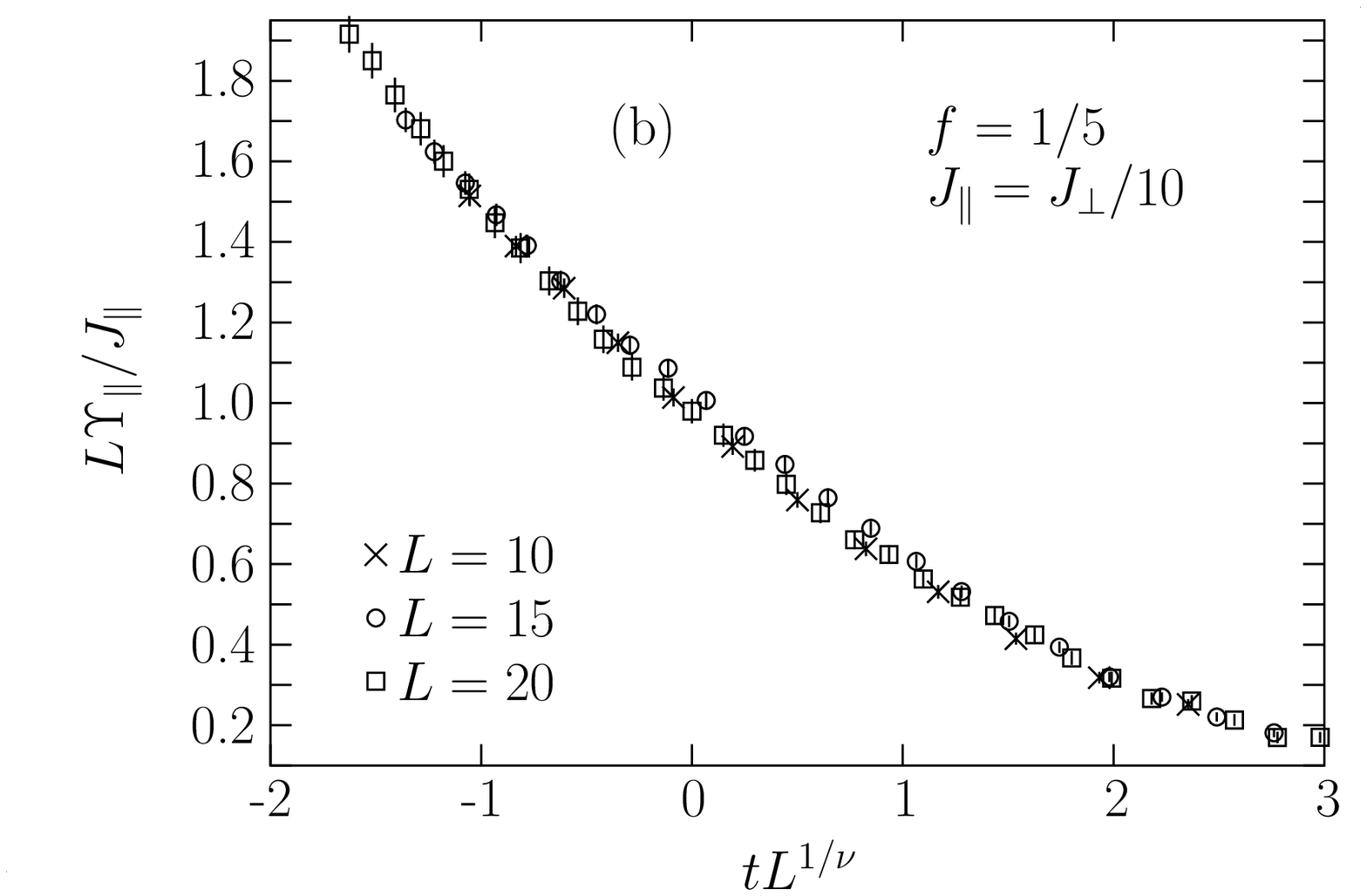}
  \caption{Data collapse for $f=1/5$ and $\Jpara/\Jperp = 1/10$. The collapse is
    done with the same $T_c=0.24$ and $\nu=1.6$ for both data sets to
    demonstrate that both quantities may be collapsed with the same parameters.}
  \label{fig:j100-perp}
\end{figure}

The simulations discussed in the previous section are for a rather strong
anisotropy, $\Jpara/\Jperp = 1/40$. It is generally expected that the critical
behavior should be independent of details as the anisotropy, and we now check
this expectation with simulations for $\Jpara/\Jperp=1/10$; data sets B and C in
\Tabsim.  \Figure{fig:j100-perp} shows scaling collapses of the helicity modulii
for data set B. Beside the weaker anisotropy the simulations also differ in that
the disorder $\varepsilon_{i\mu}$ is stronger and is now chosen from a uniform
rectangular distribution between $-1$ and 1, corresponding to $p =
\sqrt{\langle\varepsilon_{ij}^2\rangle}=1/\sqrt{3} \approx 0.577$.  In a fit
with $T_c$ and $\nu$ as adjustable parameters a collapse of $L\Uperp$ gives $T_c
= 0.239$ and $\nu=1.56$ whereas a collapse of $L\Upara$ gives $T_c = 0.241$ and
$\nu = 1.97$. The different values of $\nu$ is an indication of the rather low
precision in these determinations. In \Fig{fig:j100-perp} we show that it is
possible to collapse both sets of data with the same parameters, $T_c=0.24$ and
$\nu=1.6$. The collapse of $L\Uperp$ is very nice whereas the collapse of
$L\Upara$, especially in a region around $T_c$, is somewhat worse. However,
considering the statistical errors, we believe this to be just a statistical
fluctuation. The fact that several points around $T_c$ all deviate in the same
way is an artifact of the exchange Monte Carlo method since the exchange steps
have the effect to give correlations between results at neighboring
temperatures.

\begin{figure}[htbp]
  \includegraphics[width=7cm]{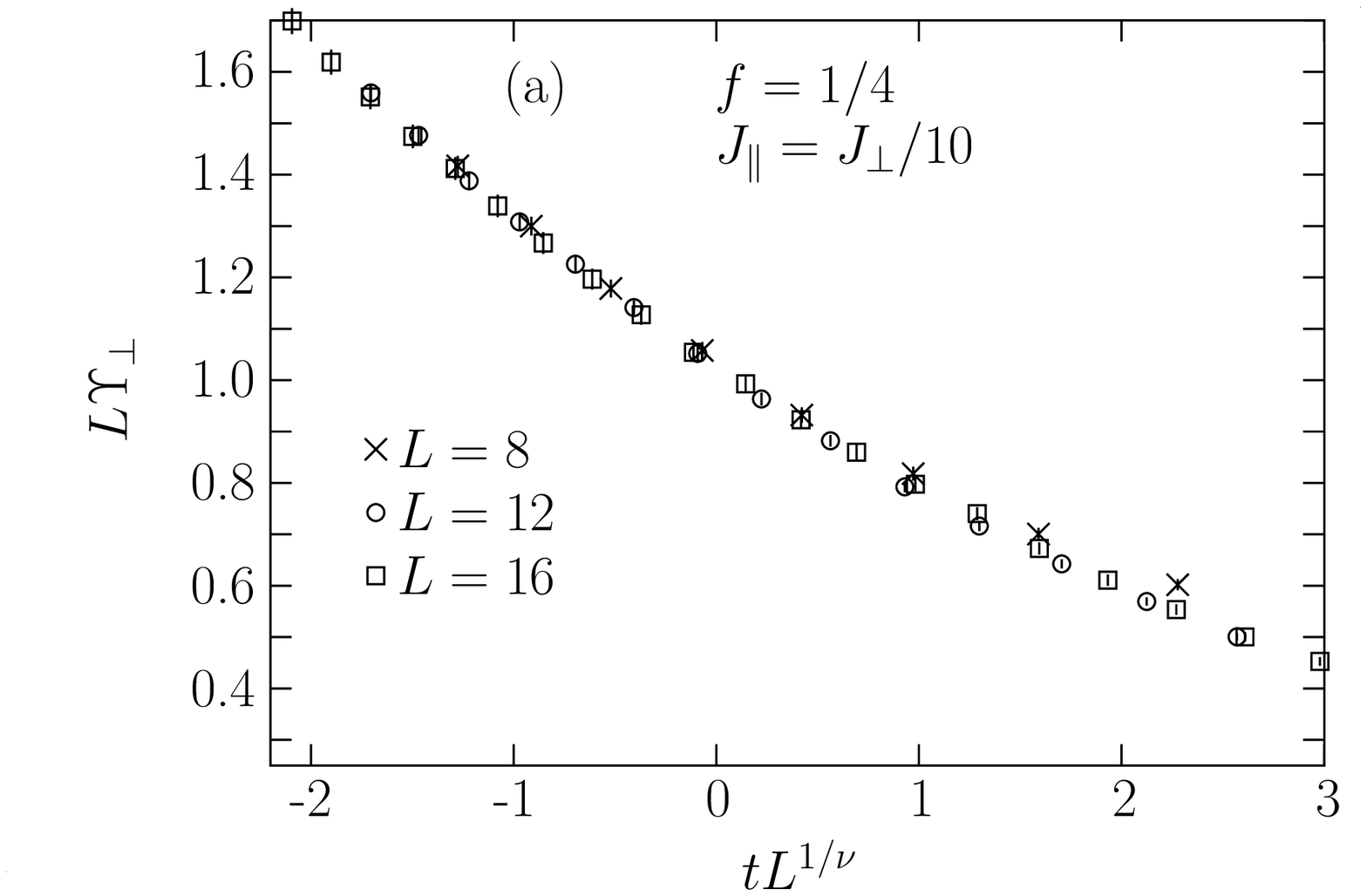}
  \includegraphics[width=7cm]{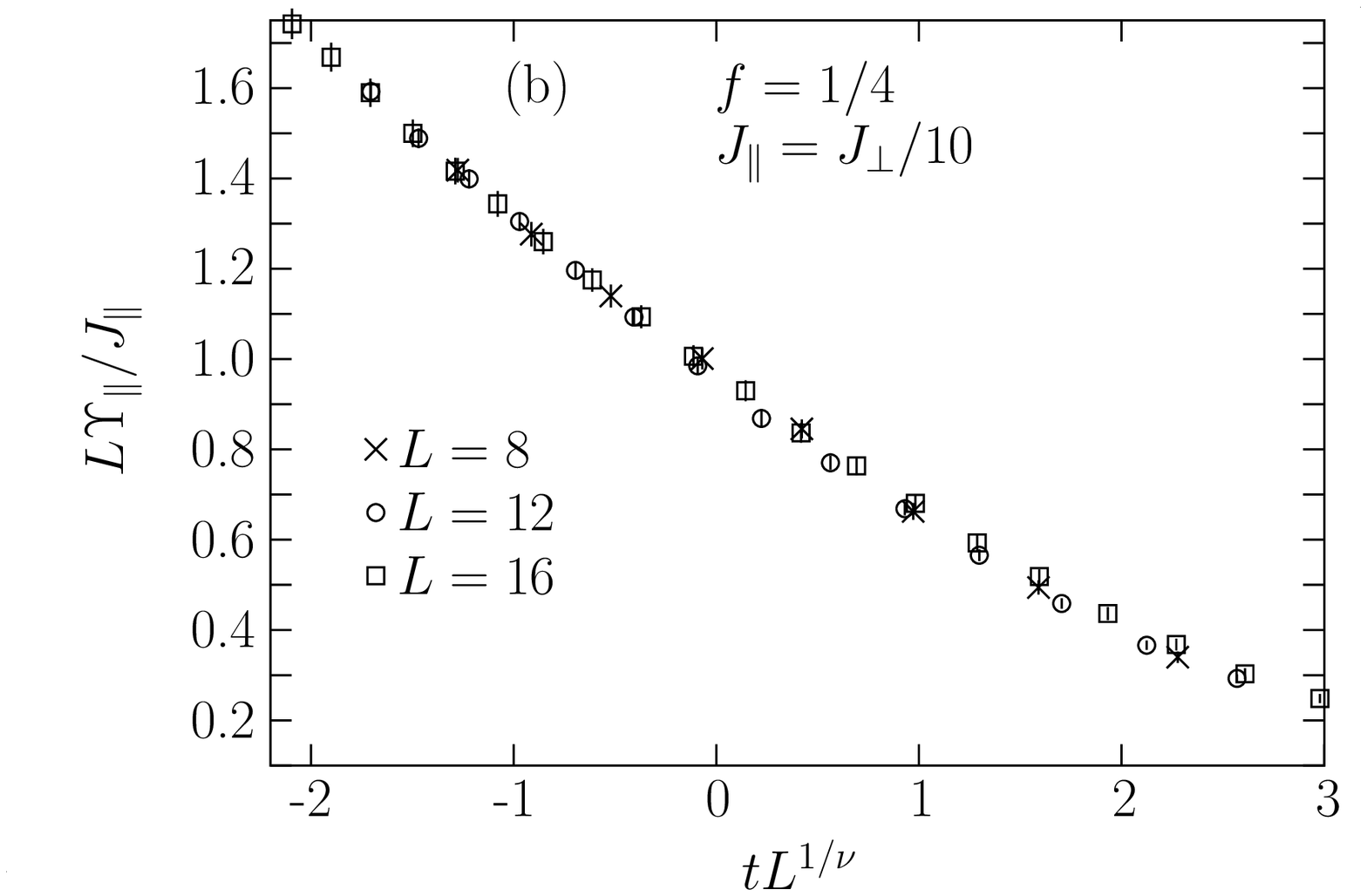}
  \caption{Data collapse of $L\Uperp$ and $L\Upara$ for $f=1/4$ and
  $\Jpara/\Jperp = 0.1$. The parameters used in the data collapses are $\nu=1.4$
  and $T_c=0.225$.}
  \label{fig:f04-j100-coll}
\end{figure}

We have also simulated the same model but with filling factor $f=1/4$; data set
C in \Tabsim. The collapse which is found in \Fig{fig:f04-j100-coll} is
excellent and we obtain $\nu = 1.35$ and $\nu=1.48$ from the scaling collapses
of $L\Uperp$ and $L\Upara$, respectively.

\subsection{Isotropic system}
\label{sec:results-iso}

The values for $\nu$ given above, obtained from simulations with different
values of anisotropy and filling factor, are within reasonable error bars
consistent with 3D gauge glass universality, $\nu\approx1.39$. This seems to
rule out the possibility that the nice scaling in \VG\ was only a coincidence.
Still, the results presented in the present section show that scaling fails when
the analysis is applied to an isotropic model.  This finding is of some
importance since isotropic couplings have been used in several
investigations\cite{Kawamura:00,Vestergren_Lidmar_Wallin,Kawamura:03} of vortex
glass models. These papers reach differing conclusions and yet other
investigations fail to find acceptable finite size scaling (private discussion).
We believe that an understanding of the problem to scale our data from isotropic
couplings may shed light on problems in these other investigations.

\subsubsection{Failure to scale the data}

For simulations of an isotropic system we use the same parameters as Kawamura in
\onlineRef{Kawamura:03} but the analysis differs from theirs in that we focus on
the behavior of the helicity modulii instead of the rms-current. 

\begin{figure}[htbp]
  \includegraphics[width=8cm]{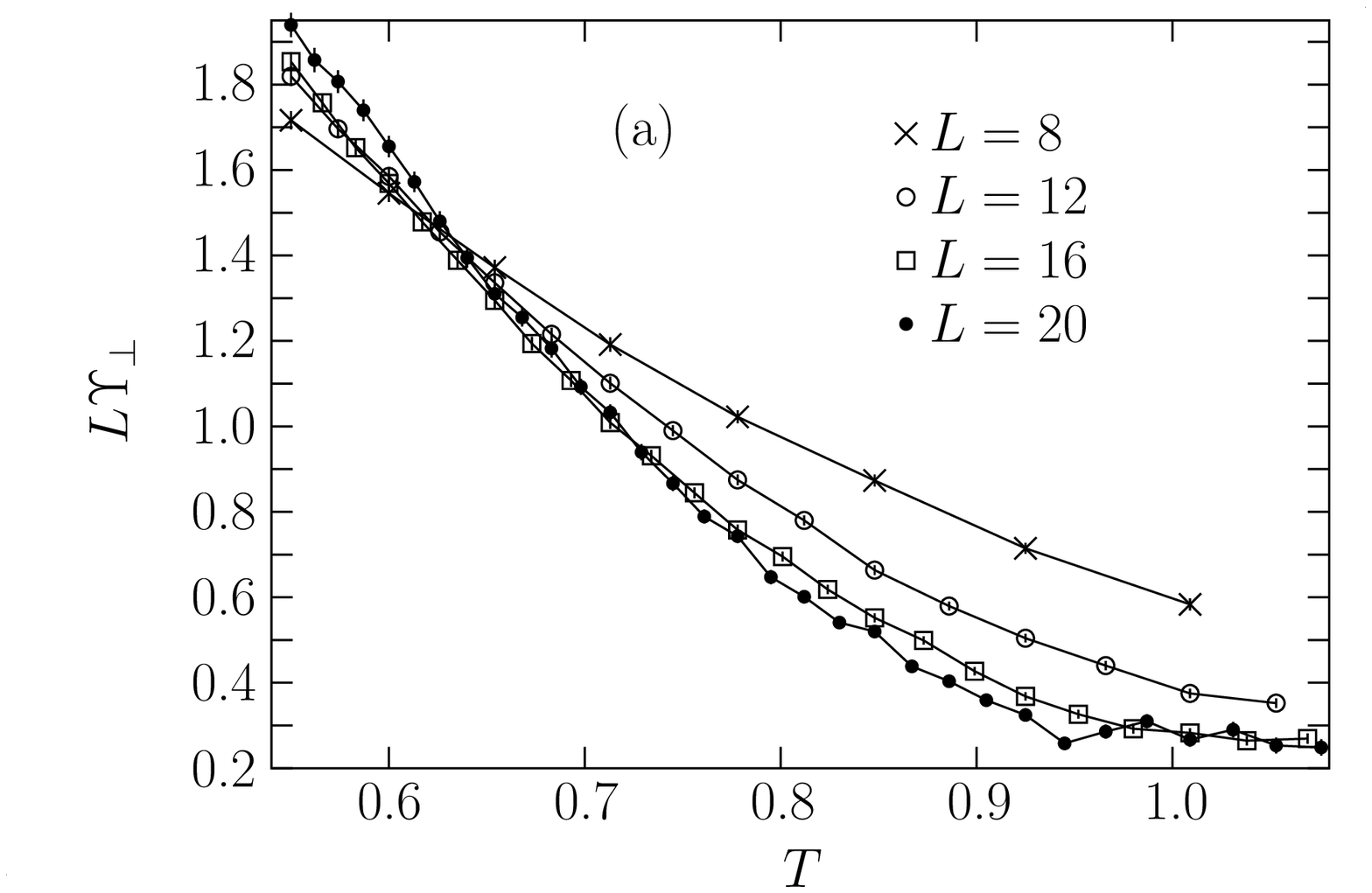}
  \includegraphics[width=8cm]{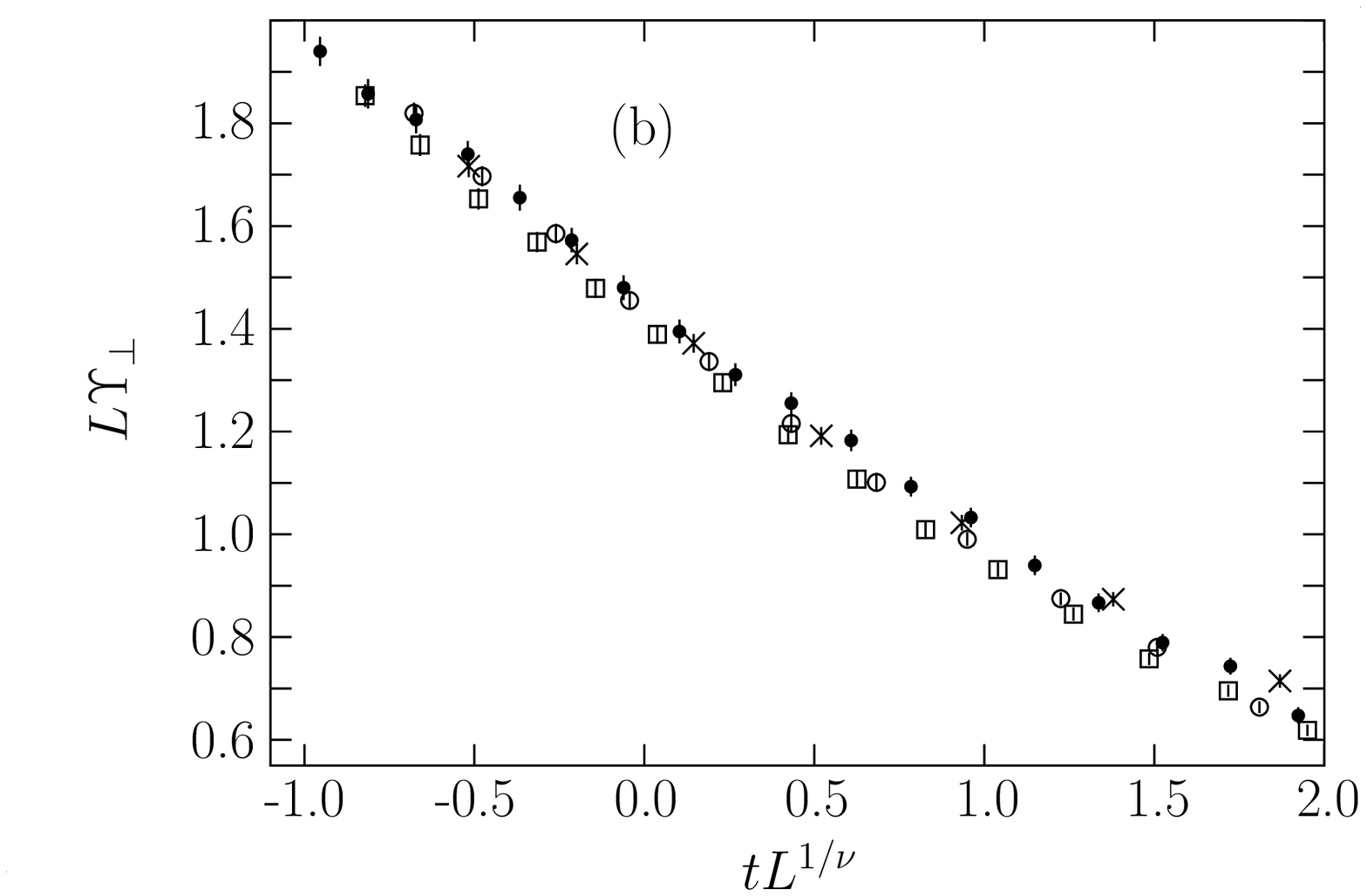}
  \caption{Raw data and attempted data collapse of $L\Uperp$ for the isotropic
  system with $f=1/4$. The parameters in the data collapse in panel (b) are
  $\nu=1.50$ and $T_c=0.63$. The value of the exponent is consistent with 3D
  gauge glass universality, but the quality of the collapse is not
  satisfactory.}
  \label{fig:f04-Uperp}
\end{figure}

\begin{figure}[htbp]
  \includegraphics[width=8cm]{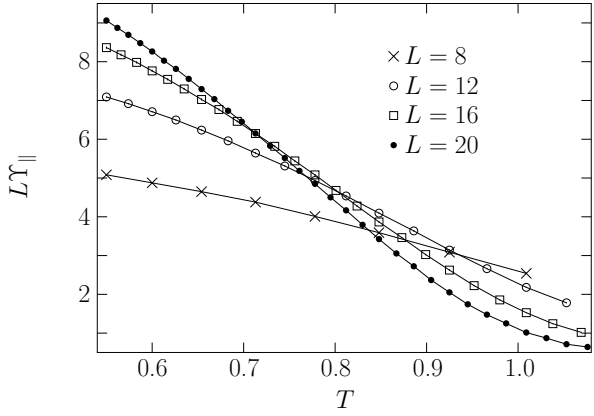}
  \caption{Raw data of $L\Upara$ for the isotropic
    system with $f=1/4$. Since the crossing temperature for two successive sizes
    shift systematically with increasing $L$ it is impossible to collapse the
    data. The large magnitude of $L\Upara$ gives additional evidence that the
    isotropic system is different from the anisotropic ones.}
  \label{fig:f04-Upara}
\end{figure}

\Figure{fig:f04-Uperp}(a) shows $L\Uperp$ for the four system sizes $L=8$, 12,
16, and 20. The data for $L\Uperp$ weakly suggests the possibility of scaling
and panel (b) shows the attempted scaling collapse with $T_c\approx 0.63$ and
$\nu\approx1.50$.  Even though the value of $\nu$ is in good agreement with our
earlier findings, the poor quality of the collapse makes it impossible to draw
any more definite conclusions.  Turning to $L\Upara/\Jpara$ shown in
\Fig{fig:f04-Upara}, we find that it is impossible to collapse the data since
the crossing points for two successive system sizes shift systematically to
lower temperatures for increasing $L$.  Beside the failure to scale the data it
should be noted that $L\Upara/\Jpara$ for the isotropic case is exceptionally
large.  For all the other cases we had $L\Upara/\Jpara\approx 1.0$ at $T_c$, but
in the isotropic model, this quantity is considerably larger in the temperature
region of interest.
  
\subsubsection{The reason for the failure to scale}
\label{sec:results-iso-reason}

The behavior of $\Upara$ in the isotropic system is thus clearly different from
the anisotropic systems with $\Jpara/\Jperp = 1/40$ or $1/10$. We will now argue
that this is because the disorder in the coupling constants is not effective in
fully disordering the system for the accessible system sizes.

\begin{figure}[htbp]
  \includegraphics[width=8cm]{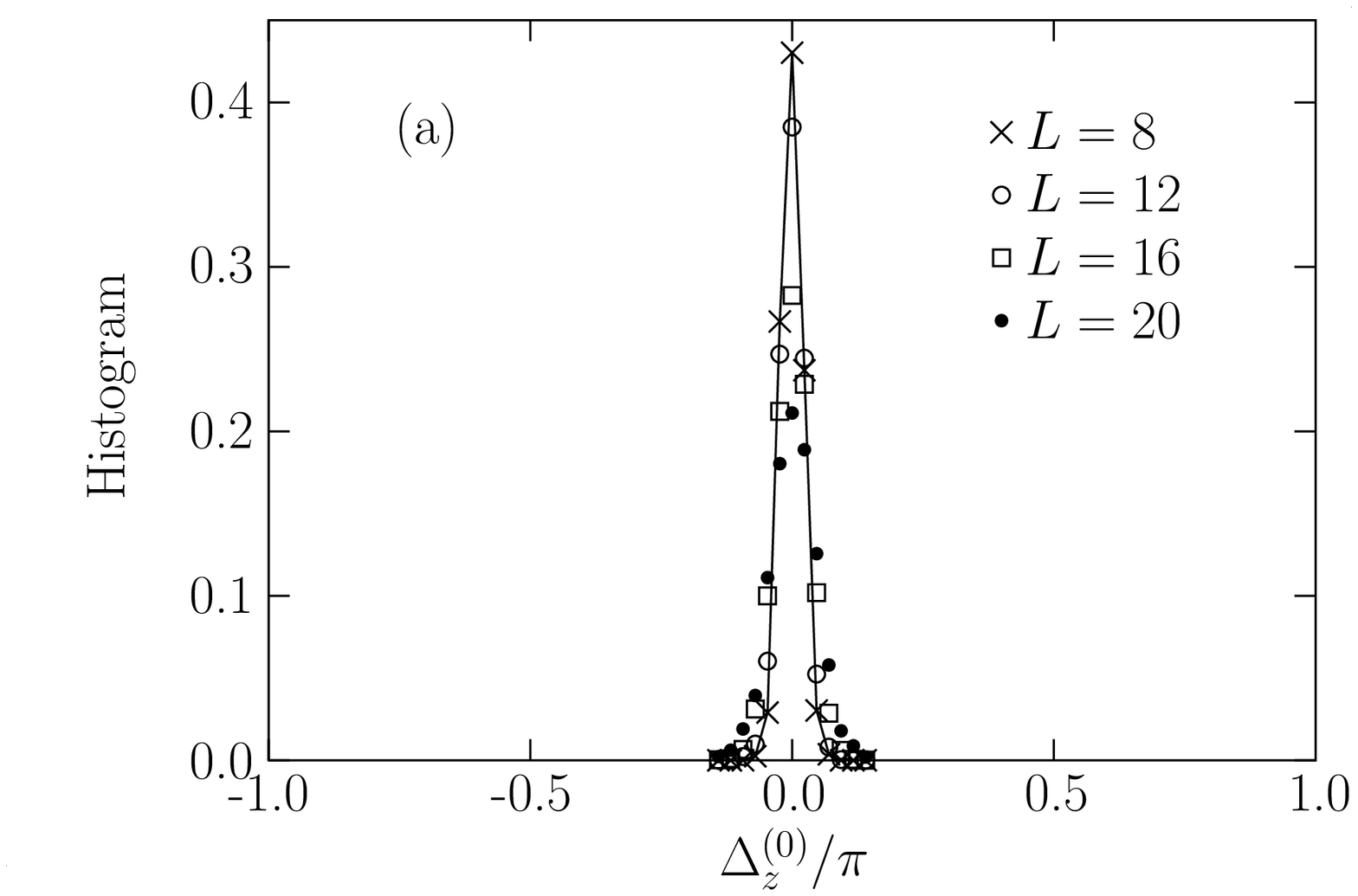}
  \includegraphics[width=8cm]{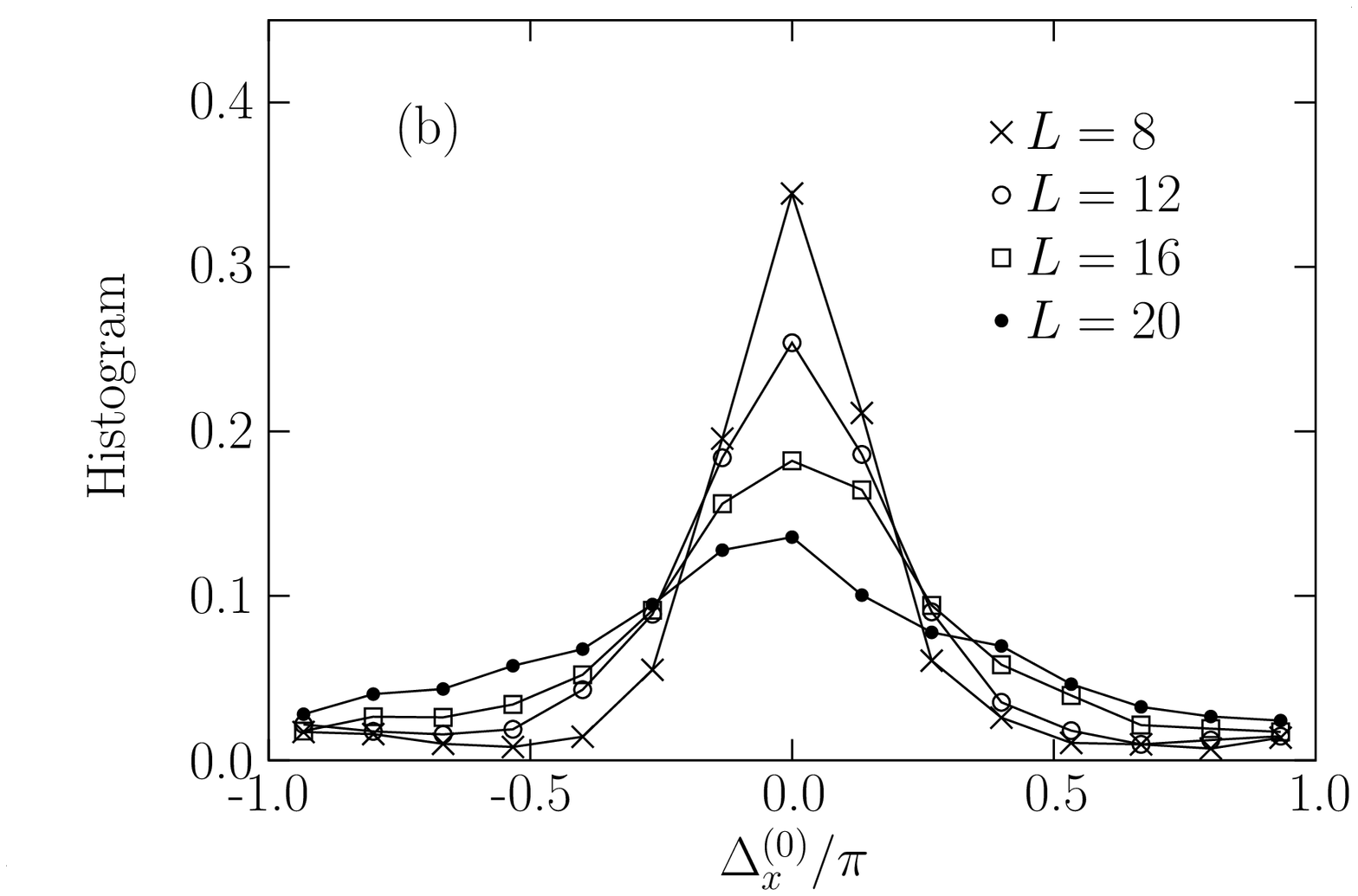}
  \caption{Histograms of $\Delta_z^{0}$ and $\Delta_x^{0}$ for the isotropic
    system with $f=1/4$.  In a fully disordered system one expects
    $\Delta_\mu^{0}$ to be uniformly distributed between $-\pi$ and $\pi$,
    but the figure shows that that is not the case for $\Jpara=\Jperp$. The peak
    around zero is strongest for $\Delta_z^{0}$ in panel (a) but is also very
    clear for $\Delta_x^{0}$ in panel (b). The histograms are calculated on
    the basis of data for all the simulated temperatures.}
  \label{postw-iso}
\end{figure}

\begin{figure}[htbp]
  \includegraphics[width=8cm]{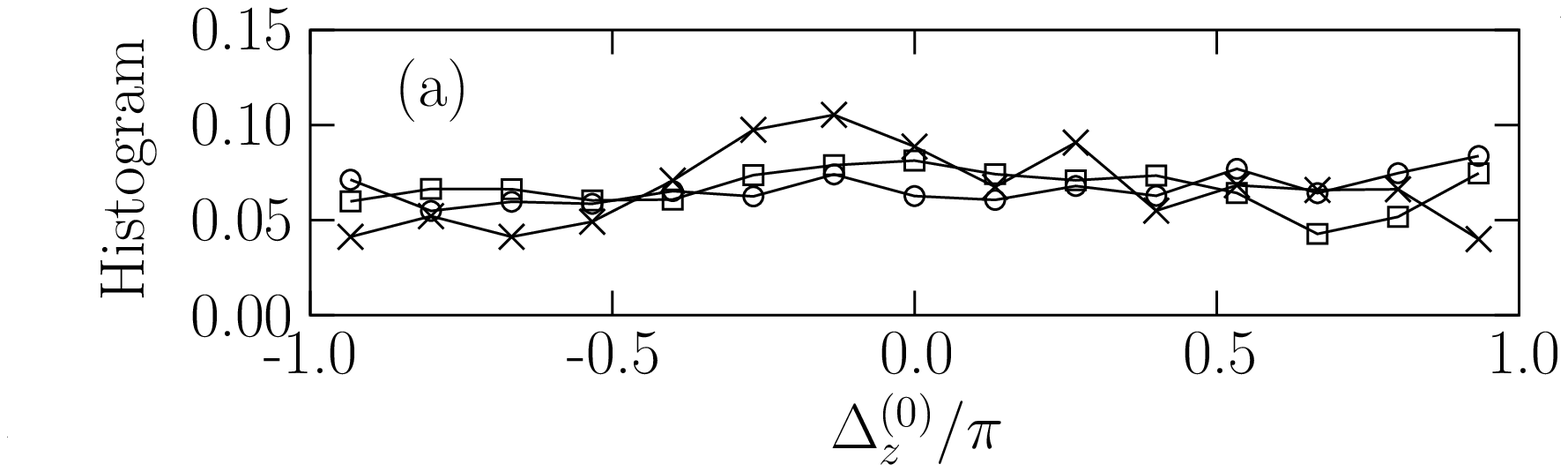}
  \includegraphics[width=8cm]{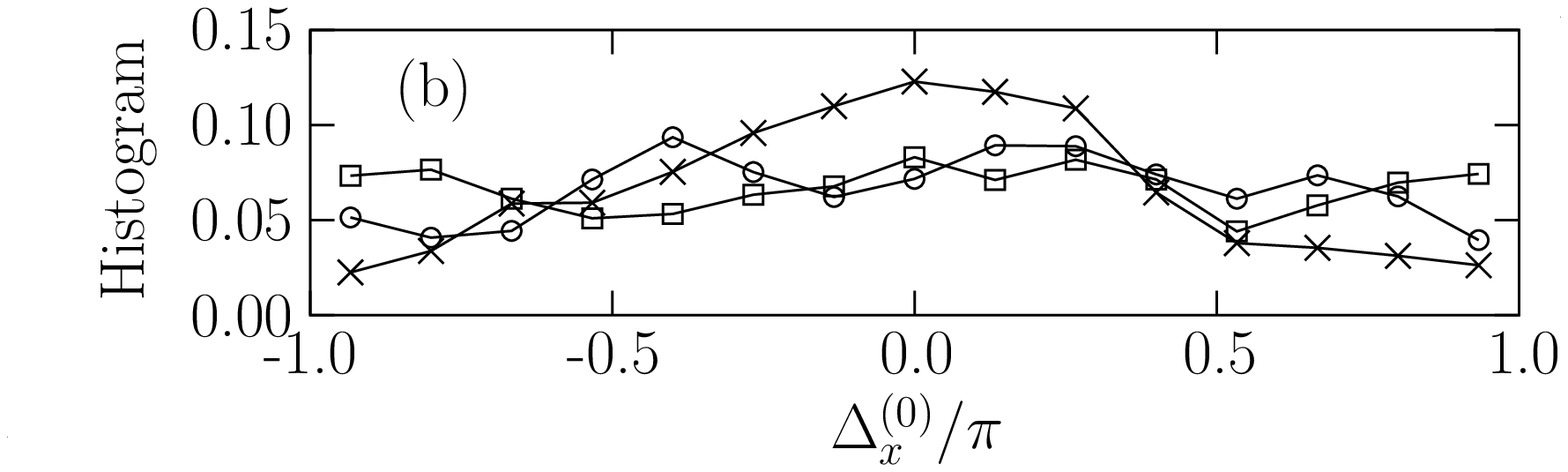}
  \caption{The figures show histograms of $\Delta_z^{0}$ and
    $\Delta_x^{0}$ for anisotropy $\Jpara/\Jperp=1/10$ and $f=1/4$ in panel
    (a) and panel (b), respectively. The results for the larger sizes, $L=12$
    and 16 (circles and squares), are consistent with a uniform distribution
    whereas the distributions for $L=8$ (crosses) have somewhat more weight
    around zero.  However, it seems that such small deviations from perfect
    disorder (a flat histogram) have no discernible effects on the scaling shown
    in \Fig{fig:f04-j100-coll}. The histograms are calculated on the basis of
    data for all the simulated temperatures.}
  \label{postw-aniso}
\end{figure}

As a probe of the loss of order we use $\Delta_\mu^{0}$ which is the position of
the minimum of the free energy, $F_\mu(\Delta_\mu)$. This quantity has been used
before as a measure of the effective strength of the
disorder.\cite{Kosterlitz_Simkin} The disorder fixed point was there
characterized by $\langle|\Delta_\mu^{0}|\rangle = \pi/2$ which corresponds to a
uniform distribution between $-\pi$ and $\pi$.  \Figure{postw-iso} shows
histograms of $\Delta_z^{0}$ and $\Delta_x^{0}$ from our data and it is clear
that the histograms are very different from a uniform distribution.  Especially
the histograms of $\Delta_z^{0}$ are very narrow with $|\Delta_z^{0}/\pi|<0.1$
for almost 99\% of the disorder realizations. For $\Delta_x^{0}$ the
distributions are considerably wider but are still clearly peaked around zero.
In both cases there is some finite size dependence, with a wider distribution
for larger system sizes.  For comparison we also show the corresponding
histograms for the anisotropic model with $\Jpara/\Jperp=1/10$ in
\Fig{postw-aniso}.  For the anisotropic case the histograms of
$\Delta_{x,z}^{0}$ are close to a uniform distribution; only the data for $L=8$
have somewhat more weight around zero. This shows that the data that exhibits
good scaling are from strongly disordered systems. In contrast, the isotropic
model appears to be far from the disorder fixed point and we believe that this
is at the root of the failure to find a convincing data collapse.

To discuss the physical meaning of $\Delta_\mu^{0}$ we return to \Fig{pair-ftbc}
which illustrates the relation between the size of a vortex pair and the value
of the twist variable in the direction perpendicular to the separation. As the
pair separates in the $x$ direction the twist $\Delta_y$ gradually increases. At
zero temperature the twist is to a good approximation proportional to the
distance, $d$, between the vortices, $\Delta_y = 2\pi d/L$.  For the more
general situation with several vortices the vortex separation generalizes to the
total dipole moment of the system of vortices, $p_x = \sum_i x_i q_i$, where $i$
enumerates the vortices, $x_i$ is the $x$- coordinate of vortex $i$, and $q_i$
is the vorticity (charge). At non-zero $T$ the distribution of $\Delta_y$ at
constant $p_x$ will be wider; the relevant expressions are given in
Ref.~\onlinecite{Gupta_Teitel_Gingras}.  For the three-dimensional case the
dipole moment generalizes to the projection of the vortex loops on a certain
plane, $C^{xy}$.\cite{Bokil_Young} The corresponding relation is then $\Delta_z
= 2\pi C^{xy}/L^2$.

In a pure system the twist histogram will always be symmetric around zero,
$\Delta_\mu^{0} = 0$, but the effect of the disorder is to favor certain vortex
loops between the layers and suppress others. The net effect may be a non-zero
$C^{xy}$ and accordingly a shift of $\Delta_z^{0}$ away from zero.  Our
interpretation of the results in \Fig{postw-iso} is therefore that the disorder
is not strong enough to introduce loops between the layers.  Note that
field-induced vortex lines that have a non-vanishing projection on the $x$-$y$
plane also contribute to $C^{xy}$.  The absence of large disorder-introduced
vortex loops between the layers (or the equivalent deflection of the
field-induced vortex lines) means that $\Delta_z^{0}$ is always close to zero.

The analyses above suggests that a strong coupling in the field direction has
the effect to reduce the amount of disorder-induced vortex loops between the
planes. The effect is to get $\Delta^0$ close to zero which means that the
effective disorder is small in the system and we believe that this is
responsible for the failure of the helicity modulii to scale. Considering the
broadening of the histograms with increasing $L$ in \Fig{postw-iso}, we expect
this to be a finite size effect, but are presently unable to estimate the size
where scaling could be expected to set in.

\section{Discussion}
\label{sec:discuss}

The use of an anisotropic model in the study of critical phenomena with finite
size scaling deserves some comments. To get data with high precision for finite
size scaling from Monte Carlo simulations, the correlation volume should ideally
have the same shape as the simulation cell. In an isotropic model a cubic
simulation cell is therefore the best choice and in the general case one wants a
common value of the fraction $\xi_\mu/L_\mu$ in all directions. For the model in
the present paper with a symmetry breaking field there is nothing that
guarantees that isotropic couplings are best. It is however possible to extract
some information about the correlations from the helicity modulii. With
$\Upsilon_\mu$ as the measure of the phase coherence, a larger $\Upsilon_\mu$
implies stronger correlations and thereby a larger correlation length in the
$\mu$ direction. By comparing data for the isotropic model in
\Figs{fig:f04-Uperp}{fig:f04-Upara}, the fact that $\Upara$ is considerably
larger than $\Uperp$ leads us to conclude that the correlations are considerably
stronger in the field direction compared to the perpendicular direction. One way
to reach the goal of a simulation cell with the same shape as the correlation
volume would then be to increase the aspect ratio $L_z/L$, but a different and
more efficient way is to instead decrease $\Jpara$, the coupling strength in the
field direction.

To get a better understanding of the effect of anisotropic couplings on the
helicity modulii we have made some additional simulations on the ordinary 3D XY
model (zero field and no disorder) with $J_z/J = 1/4$. Since one expects
$\xi_\mu\propto \sqrt{J_\mu}$ the aspect ratio was then chosen to be $L_z/L=1/2$
which gives a simulation cell with the same shape as the correlation volume.
With this value of the aspect ratio the simulations give $\Upsilon/J =
\Upsilon_z/J_z$ at $T_c$ to a good approximation. As shown in
\Fig{fig:j100-perp} the same relation holds to a good approximation at and close
to $T_c$ in the simulations of the vortex glass with $\Jpara/\Jperp = 1/10$ and
$L_z/L=1$. This suggests that the correlations in the different directions are
about equally strong when the anisotropy is set to $\Jpara/\Jperp = 1/10$ and
that this value therefore is close to optimal for the anisotropy in the vortex
glass simulations with $f=1/5$.

Even though it thus seems that our model is best examined with a rather large
anisotropy we now turn to the results obtained with isotropic couplings. These
simulations were performed with the parameters of \onlineRef{Kawamura:03} to
make it possible to directly compare the results. It is however clear that the
results are significantly different.  Whereas our $L\Uperp$ almost collapse at
$T=0.63$ with $\nu\approx 1.5$ their corresponding quantity, $I_T$, collapses
for the three largest sizes at $T_g=0.81$ with $\nu= 1.0$. Especially the
different values of the critical temperature points to a systematic difference.

We believe that the reason for this difference is their calculation of $\Irms$
as the derivative of $F(\Delta)$ evaluated at $\Delta=0$, rather than at random
values of $\Delta$. From our direct determinations of $F(\Delta_\mu)$ we have
found that the typical structure of this quantity (obtained with parameter set
D) is a single minimum of the free energy with a shape that in most cases to a
very good approximation is parabolic, $F(\Delta_\mu) = \mathrm{const} +
\Upsilon_\mu (\Delta_\mu - \Delta_\mu^0)^2/2$, where both $\Upsilon_\mu$ and
$\Delta_\mu^0$ depend on the disorder realization. When the derivative is
evaluated at $\Delta_\mu=0$ one gets $I_\mu = -\Upsilon_\mu \Delta_\mu^0$. This
means that the obtained rms-current is not only a measure of the amount of
structure in $F(\Delta_\mu)$ but also depends on the location of this structure.
Against that backgound the size-dependence of the distribution of $\Delta_\mu^0$
shown in \Fig{postw-iso} is problematic and we believe this to be the reason for
the different critical behavior in \onlineRef{Kawamura:03} compared to the
results in this paper.  This undesired finite size will affect the determination
of the rms-current and we therefore believe that the scaling behavior of $I_T$
in \onlineRef{Kawamura:03} is only accidental.

The failure of the helicity modulus to scale in the isotropic model is a related
but different question. As discussed in Sec.~\ref{sec:results-iso-reason} a
message from \Fig{postw-iso} is that the isotropic model is not sufficiently
disordered for the simulated system sizes and it seems possible that this
remaining order destroys the transition. The broadening of the histograms in
\Fig{postw-iso} with increasing system size would lead to a flat distribution in
the limit of large $L$ and one would then expect scaling with the 3D gauge glass
exponents.  However, considering the slow widening of the histograms as $L$
increases, scaling would presumably only be seen for very large systems.

It should finally be noted that the problem with determining the rms-current
from the derivative at $\Delta=0$ is not present in the 3D gauge glass model.
The reason is that the randomness there is put into the vector potential and
that a random $\Delta$ then may be absorbed in the similarly random $A_{ij}$.
There is then no need to make use of a random $\Delta$ and the standard way to
evaluate $\Irms$ at $\Delta=0$ is acceptable.

To summarize: the main conclusion of the present investigation is that the
vortex glass model is in the same universality class as the 3D gauge glass
model. This is a confirmation of the behavior found in \VG. Still, it is found
that simulations with isotropic couplings do not give any convincing scaling
collapse and we argue that the reason is that the effective randomness for the
accessible system sizes is too small to give the correct behavior of the vortex
glass transition.

\section{Acknowledgements}

We acknowledge discussions with S. Teitel. This work was supported by the
Swedish Research Council, contract No.\ 2002-3975, and by the resources
of the Swedish High Performance Computing Center North (HPC2N).


\begin{thebibliography}{28}
\bibitem[{why()}]{why-this-delta}
\bibinfo{note}{The splitting up of $\Delta_\mu$ in $L_\mu$ different variables
  has been done for technical reasons. With the angles $\theta_i$ discretized
  to 250 different values (divisible by $1/f=5$) the minimum change of
  $\delta_\mu$ (as a single variable) would for a system with $L_\mu=25$ give
  changes of $\Delta_\mu$ as large as 25/250. This would lead to only ten
  different values of $\Delta_\mu$. By instead using $L_\mu$ different
  variables $\delta_\mu^{(\mathbf{r}_i\cdot \hat{\mu})}$ the sum $\Delta_\mu$
  may take on all the possible values.}
\end{thebibliography}

\end{document}